\documentclass[12pt,preprint]{aastex}
\usepackage{lscape}

\slugcomment{\it Accepted for Publication by the Astronomical Journal}
\shortauthors{Yan et al.}
\shorttitle{Characterizing \wise\, Extragalactic Sources}

\def\ie{{i.e.,}}
\def\eg{{e.g.,}}

\def\deg{\ifmmode {^{\circ}}\else {$^\circ$}\fi}
\def\kms{\ifmmode {\rm\,km\,s^{-1}}\else
    ${\rm\,km\,s^{-1}}$\fi}
\def\ergcm2s{\ifmmode {\rm\,ergs\,cm^{-2}\,s^{-1}}\else
    ${\rm\,ergs\,cm^{-2}\,s^{-1}}$\fi}
\def\ergAcm2s{\ifmmode {\rm\,ergs\,cm^{-2}\,s^{-1}\,\AA^{-1}}\else
    ${\rm\,ergs\,cm^{-2}\,s^{-1}\,\AA^{-1}}$\fi}
\def\ergs{\ifmmode {\rm\,ergs\,s^{-1}}\else
    ${\rm\,ergs\,s^{-1}}$\fi}
\def\kmsMpc{\ifmmode {\rm\,km\,s^{-1}\,Mpc^{-1}}\else
    ${\rm\,km\,s^{-1}\,Mpc^{-1}}$\fi}

\def\lya{Ly$\alpha$}

\def\nev{\ion{Ne}{5} $\lambda$3426}
\def\civ{\ion{C}{4} $\lambda$1549}
\def\ciii{\ion{C}{3}] $\lambda$ 1909}
\def\oii{[\ion{O}{2}] $\lambda$3727}
\def\oiii{[\ion{O}{3}] $\lambda$5007}

\def\mgii{[\ion{Mg}{2}] $\lambda$2800}

\def\spose#1{\hbox to 0pt{#1\hss}}
\def\simlt{\mathrel{\spose{\lower 3pt\hbox{$\mathchar"218$}}
     \raise 2.0pt\hbox{$\mathchar"13C$}}}
\def\simgt{\mathrel{\spose{\lower 3pt\hbox{$\mathchar"218$}}
     \raise 2.0pt\hbox{$\mathchar"13E$}}}

\def\gs{\mathrel{\raise0.35ex\hbox{$\scriptstyle >$}\kern-0.6em
\lower0.40ex\hbox{{$\scriptstyle \sim$}}}}
\def\ls{\mathrel{\raise0.35ex\hbox{$\scriptstyle <$}\kern-0.6em
\lower0.40ex\hbox{{$\scriptstyle \sim$}}}}

\newcommand{\um}{\,$\mu$m}

\newcommand{\wise}{{\it WISE}}
\newcommand{\iras}{{\it IRAS}}

\newcommand{\spitzer}{{\it Spitzer}}
\newcommand{\herschel}{{\it Herschel}}

\begin{document}

%\title{What does the Wide-Field Infrared Survey Explorer (WISE) Reveal ? \\
\title{Characterizing the Mid-IR Extragalactic Sky with WISE and SDSS} 
%Revealed by the Wide-Field Infrared Survey Explorer}

\author{Lin Yan$^1$, E. Donoso$^1$, Chao-Wei Tsai$^1$, D. Stern$^2$, R. J. Assef$^{2,8}$, P. Eisenhardt$^2$, A. W. Blain$^3$, R. Cutri$^1$, T. Jarrett$^1$, S. A. Stanford$^4$, E. Wright$^5$, C. Bridge$^6$, D.A. Riechers$^{6,7}$}

\affil{$^1$Infrared Processing and Analysis Center, California
Institute of Technology, MS 100-22, Pasadena, CA 91125, USA}

\affil{$^2$Jet Propulsion Laboratory, California Institute of
Technology, Pasadena, CA 91109, USA}

\affil{$^3$Department of Physics \&\ Astronomy, University of Leicester, University Road, Leicester, LE1 7RH, UK}

\affil{$^4$Department of Physics, University of California, Davis,
CA 95616, USA}

\affil{$^5$Astronomy Department, University of California, Los
Angeles, CA 90095-1547, USA}
\affil{$^6$Astronomy Department, California Institute of Technology, 1200 East California Blvd, Pasadena, CA91125, USA}

\affil{$^7$Astronomy Department, Cornell University, Ithaca, NY 14853, USA}
\affil{$^8$NASA Postdoctoral Program}

\email{lyan@ipac.caltech.edu}

\begin{abstract}

The {\it Wide-field Infrared Survey Explorer (WISE)} has completed
its all-sky survey in four channels at 3.4\,-\,22\um, detecting
hundreds of millions of objects.  We merge the \wise\, mid-infrared
data with optical data from the Sloan Digital Sky Survey (SDSS) and
provide a phenomenological characterization of \wise\, extragalactic
sources.  \wise\, is most sensitive at 3.4\um\ ($W1$) and least
sensitive at 22\um\ ($W4$).  The $W1$ band probes massive early-type
galaxies out to $z \simgt 1$.  This is more distant than SDSS identified
early-type galaxies, consistent with the fact that 28\%\ of 3.4\um\
sources have faint or no $r$-band counterparts ($r > 22.2$).  In
contrast, 92 - 95\%\ of 12\um\ and 22\um\ sources have SDSS optical
counterparts with $r \le 22.2$.  \wise\ 3.4\um\ detects 89.8\%\ of the entire
SDSS QSO catalog at SNR$_{W1}$\,$>$7$\sigma$, but only 18.9\%\ at 22\um\ with
SNR$_{W4}$\,$>$5$\sigma$. We show that \wise\, colors alone
are effective in isolating stars (or local early-type
galaxies), star-forming galaxies and strong AGN/QSOs at $z \simlt
3$.   We highlight three major applications of \wise\, colors: (1) Selection of strong AGN/QSOs at $z \le 3$ using
$W1 - W2 > 0.8$ and $ W2 < 15.2$ criteria, producing a better census
of this population.  The surface density of these strong AGN/QSO candidates is
$67.5\pm0.14$ per deg$^2$.  (2) Selection of dust-obscured, type-2
AGN/QSO candidates. We show that \wise\, $W1 - W2 > 0.8$, $W2<15.2$ combined with $r -
W2 > 6$ (Vega) colors can be used to identify type-2 AGN candidates.  
The fraction of these type-2 AGN candidates is 1/3rd of all \wise\ color-selected AGNs.
 (3) Selection of ultra-luminous
infrared galaxies at $z \sim 2$ with extremely red colors, $r - W4
> 14$ or well-detected 22\um\ sources lacking
detections in the 3.4 and 4.6\um\ bands.  The surface density of
$z \sim 2$ ULIRG candidates selected with $r - W4 > 14$ is 
$0.9\pm0.07$ per deg$^2$ at SNR$_{W4}\ge5$ (the corresponding, lowest flux density of 2.5\,mJy),
which is consistent with that inferred from smaller area \spitzer\,
surveys. Optical spectroscopy of a small number of these high-redshift ULIRG candidates
confirms our selection, and reveals a possible trend that optically fainter or 
$r - W4$ redder candidates are at higher redshifts. 

\end{abstract}


\section{Introduction}

With the power of large number statistics, wide-area astronomical
surveys in this decade have made many significant discoveries. A
new generation of high-precision, wide-area optical imaging surveys
are being proposed for the end of this decade to address fundamental
questions in cosmology, galaxy formation and to search for extrasolar
planets. Most recently, the advent of the {\it Wide-field Infrared
Survey Explorer} \citep[\wise;][]{ned2010} has provided the community
with an unprecedented dataset in the mid-infrared. This unique NASA
mission mapped the entire sky in four bands at 3.4, 4.6, 12 and
22\um\, ($W1$ through $W4$), with 5$\sigma$ point source sensitivities
better than 0.05, 0.1, 0.75 and 6 mJy, respectively.  \wise\, 12\um\
images are more than 100 times deeper than previous all-sky infrared
survey missions, such as that provided by the {\it Infrared
Astronomical Satellite} \citep[\iras;][]{gerry84}, while the 3.4\um\
data are 1.5 magnitude (a factor of 4 in flux density) deeper than the Two-Micron All-Sky
Survey \citep[2MASS;][]{skrutskie06} $K_s$ data for sources with
spectral energy distributions similar to an A0 star; \wise\, is
even more sensitive for red sources like K-stars, early-type galaxies, AGN, and
dust-obscured galaxies.

\wise\, has made two public data releases, including coadded atlas
images and source catalogs. The first  was the preliminary release
in April 2011, covering roughly half of the sky, and the second was
the all-sky data release in March 2012, covering the entire sky.
Although \wise\, has a significantly smaller aperture and the angular resolution
is only half that of \spitzer, its unique all-sky coverage enables
selection of large samples of extragalactic sources, allowing
statistical studies of stellar photospheric emission at 3.4 and
4.6\um\ and dust emission at 12 and 22\um. The primary goal of this
paper is to provide an empirical characterization of \wise\,
extragalactic sources, and to identify the types of sources which
can be isolated using mid-infrared colors.  To achieve this goal,
we combine \wise\, mid-infrared data with optical data from the
Sloan Digital Sky Survey (SDSS) seventh data release
\citep[DR7;][]{abazajian}.  We characterize the color distributions
and source types using broad-band photometry and spectroscopic
information.  The motivation for such a phenomenological study is
to give the community a summary of the observational properties and
the limitations of two large surveys, particularly for mid-infrared
extragalactic objects.  More detailed, quantitative analyses of
specific extragalactic populations are discussed in companion papers
by the \wise\, extragalactic science team.  Specifically,
\citet{griffith11} and \citet{tsai12a} discuss \wise\, selection of low-metallicity
blue compact dwarf galaxies at $z \simlt 0.1$.  \citet{emilio12}
examines the origin of 12\um\ emission in mid-infrared galaxies at
$z \sim 0.1$ using \wise\, data in conjunction with SDSS DR7
spectroscopic data, while \citet{lake12} uses Keck spectroscopy to
study the redshift distribution of flux-limited \wise\, samples.
\citet{stern12} and \citet{assef12} present detailed studies of
\wise-selected AGN within the COSMOS and Bo\"otes fields, respectively,
including careful analyses of completeness and reliability of \wise\,
AGN selection using $W1 - W2$ color.  \citet{eisenhardt12} presents
the first results for $z \sim 2$ ultraluminous infrared galaxies
(ULIRGs) discovered by \wise.  \citet{jingwen12} and \citet{yan12}
present the far-infrared properties and spectral energy distributions
(SEDs) of similarly selected galaxies based on ground-based millimeter
and \herschel\ space-based far-infrared data, respectively.
\citet{bridge12} discusses the interesting Ly$\alpha$ properties
of the \wise\, ULIRG population, emphasizing the high rate of
extended emission, so-called ``Ly$\alpha$ blobs''.  \citet{tsai12b}
and Stern et al. (in prep) present detailed studies of two interesting
\wise-selected AGN, while \citet{jarrett12} and \citet{petty12}
study spatially resolved, local ($z < 0.1$) galaxies with \wise.
\citet{Blain12} presents the \wise\, properties  of $z \ge 6$ optically
selected QSOs. Finally, \citet{gettings12} discusses the first results of using
\wise\, to identify high-redshift galaxy clusters.

The organization of this paper is as follows. Section~\ref{sec-data}
describes the \wise\, and SDSS data, including the \wise\, sample
selection and the SDSS optical data used by this paper.
Section~\ref{sec-results} presents the main results, providing the
technical details of the catalog matching and summarizing the
brightness, color and photometric redshift distributions for galaxies
detected by both \wise\, and SDSS (\S~3.1).  We identify the
characteristic color criteria which can be used to select large
samples of strong AGN/QSOs (\S~3.2), type-2 AGN candidates (\S~3.3)
and potential $z \sim 2$ ULIRGs (\S~3.4).  Section~\ref{sec-discuss}
summarizes the main conclusions and discusses the implications for
future studies using \wise.  Throughout the paper, we adopt an
$\Omega_M$\,=\,0.27, $\Omega_\Lambda$\,=\,0.73\ and $H_0$\,=\,71\kmsMpc\
cosmology.

%[{\it Maybe here is where to discuss AB vs. Vega magnitudes?}]

\section{Data \label{sec-data}}

\subsection{WISE-SDSS sky coverage \label{sec-cov}} 

\begin{figure}[!h]
\plotone{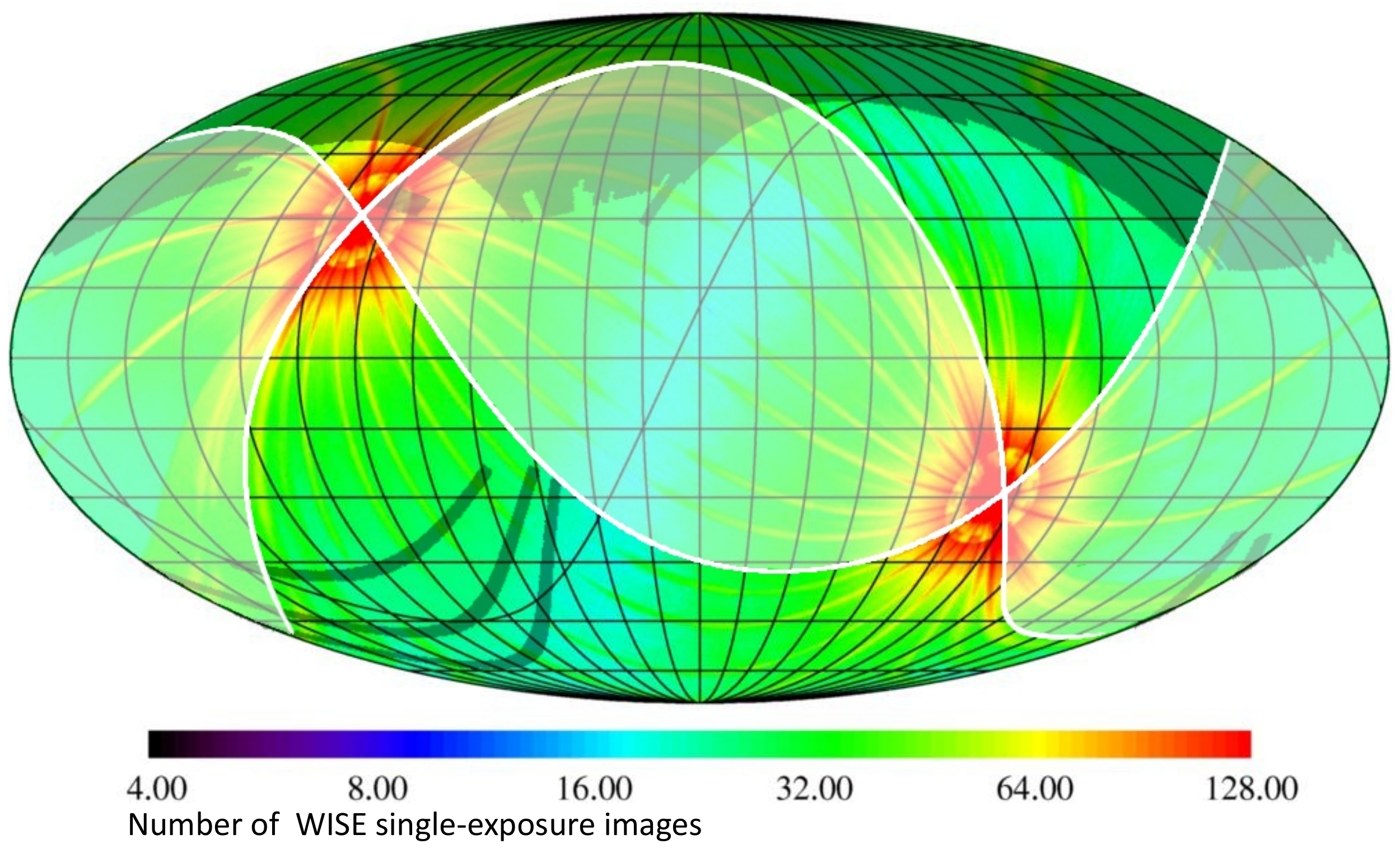}
\caption{Sky coverage of \wise\, and SDSS DR7 in Galactic Coordinates.  The background map
shows the \wise\, all-sky depth with colors indicating the number
of single-exposure frames at each sky position. The white overlaid
area shows the sky region covered by the \wise\, preliminary public
data release, covering $57$\%\ of the full sky.  The dark green
region, primarily towards the North Galactic cap, is covered by the
SDSS DR7 data. This paper analyzes data drawn from the overlap
regions between the \wise\, preliminary public data release and
SDSS DR7, a region covering 2344 deg$^2$.
\label{wcover}}
\end{figure}

The data used by this paper are drawn from the overlapping sky
region where both \wise\, preliminary public release data and SDSS
DR7 data are available. Although this area is only 1/4th of the
total SDSS area, the number of galaxies included in our analysis
is $> 10^7$, large enough for the purposes of this paper. Because this
paper started with the preliminary release data, we kept the same overlap region between
the preliminary release data and SDSS DR7, but the actual \wise\ photometry are taken from the all-sky data release.
Figure~\ref{wcover} presents the full-sky coverage map from \wise,
with colors indicating the number of repeated single exposures at
each sky position.  Overlaid in the figure are the coverages of
the SDSS DR7 data and the \wise\, preliminary release data.  The
overlap area between DR7 and the \wise\, preliminary release is
2344 deg$^2$ and defines the area analyzed
in the rest of this paper.  This area is slightly less than 30\%\
of the total SDSS areal coverage, and is 5.7\%\ of the entire sky.

One characteristic feature of the \wise\, mission is that it does
not have uniform depth-of-coverage.
For both the preliminary and the all-sky data release, the median depth-of-coverage is 
15.65, 15.55, 14.85 and 14.84 exposures at $W1,W2,W3$ and $W4$ respectively (each single exposure has 11\,seconds). 
95\%\ of the sky has coverage $\geq10.82$ at $W1$, and $\geq9.90$ at $W4$\footnote{See Section 6 in the Explanatory Supplement to the \wise\ All-Sky Data Release Products, {\it http://wise2.ipac.caltech.edu/docs/release/allsky/expsup/index.html}}. Taking 10 exposures per position as the depth-of-coverage in $W1$ and $W2$, and 9 in $W3$ and $W4$, the $5\sigma$ sensitivities are 17.30\,mag (0.037\,mJy), 15.84\,mag (0.079\,mJy), 11.59\,mag (0.67\,mJy) and 8.00\,mag (5.10\,mJy) for the four bands, respectively. 

% FIGURE 1:  areal coverage

\subsection{WISE data\label{sec-wisedata}}

We start with the \wise\, preliminary public release source
catalog \citep{cutri11}\footnote{A detailed description of the catalog can be found
in the Explanatory Supplement to the \wise\, Preliminary Data Release Products
at {\it http://wise2.ipac.caltech.edu/docs/release/prelim/expsup/.}}.
This catalog includes all sources detected in at least one band
with photometric signal-to-noise ratio (SNR)\,$\ge 7$ after removing
all artifacts.  As described in the explanatory supplement to the
preliminary data release products, source detection and photometric
measurements are performed on all four bands simultaneously.  The
released catalog contains photometric measurements (or limits) in
all four bands.

\wise\, is the most sensitive at 3.4\um\, and the least sensitive
at 22\um; the minimum $5\sigma$ sensitivities for these two bands
are 0.05 and 6\,mJy, respectively.  For
$f_\nu$\,$\propto$\,$\nu^\alpha$, the sensitivity ratio $f_\nu(3.4\mu
m)/f_\nu(22\mu m) = [\nu(3.4\mu m)/\nu(22\mu m)]^\alpha = 0.0083$
corresponds to a mid-infrared spectral index $\alpha \sim -2.56$.  Among all \wise\, $W1$-detected galaxies at the limiting
depth of that band, only 2\%\ of the sources are also detected in $W4$, implying extremely red mid-infrared spectral energy distributions (SEDs), redder than $f_\nu \propto \lambda^{2.56}$.

The public released catalog contains a very small number of
sources which are detected only in $W3$ and/or $W4$, but not in
$W1$ and/or $W2$ \citep[\eg][]{eisenhardt12}. Due to the large
sensitivity differences between $W1$ and $W4$, the entire catalog
is essentially a 3.4\um-selected sample with SNR\,$\ge$\,7 ($\sigma_{W1}
\sim 0.15$); only 0.16\%\ of the catalog is very faint in $W1$ but
bright in one of the other three \wise\, bands (\eg\, SNR$_{W1} <
7$, but SNR$>$\,7 for any of the other three bands).  Therefore,
we call this original catalog from the \wise\, data archive, without
any additional cuts, the W1 sample.  Many sources in this
preliminary public release catalog have SNR$_{W1} > 7$, but low SNR
in the other three \wise\, bands.  When we study the color
distributions, it becomes necessary to require the color errors to
be small relative to the range of colors.  Specifically, the $W1 -
W2$ color for \wise\, galaxies roughly spans the narrow range, 0 to
2.  We require the errors in $W1 - W2$ to be less than 0.27 mag,
meaning we need a source catalog with SNR$_{W2} \ge 5$ (corresponding
to $\sigma_{W2} < 0.22$ mag).  A similar argument is applied to the
$W2 - W3$ colors, which have a larger spread of $0.5 - 5$.  In this
case, we require SNR$_{W3} \ge 3$ ($\sigma_{W3} < 0.35$ mag).  In
summary, to obtain meaningful color distributions, we base our
analyses below on the following three source samples: the original
W1 sample (SNR$_{W1}\ge7$), a W1/2 sample (SNR$_{W1}\ge7$ and SNR$_{W2}\ge5$), and a W1/2/3 sample  (${\rm SNR}_{W1}\ge7,{\rm SNR}_{W2}\ge5,{\rm SNR}_{W3}\ge3$).

\begin{figure}[!h]
% \plotone{/Users/linyan1/wise-science/papers/data/plots/delivery_fig_2011_0421/Fig2.jpg}
% \plotone{/Users/linyan1/wise-science/papers/data/plots/bar2.pdf}
\plotone{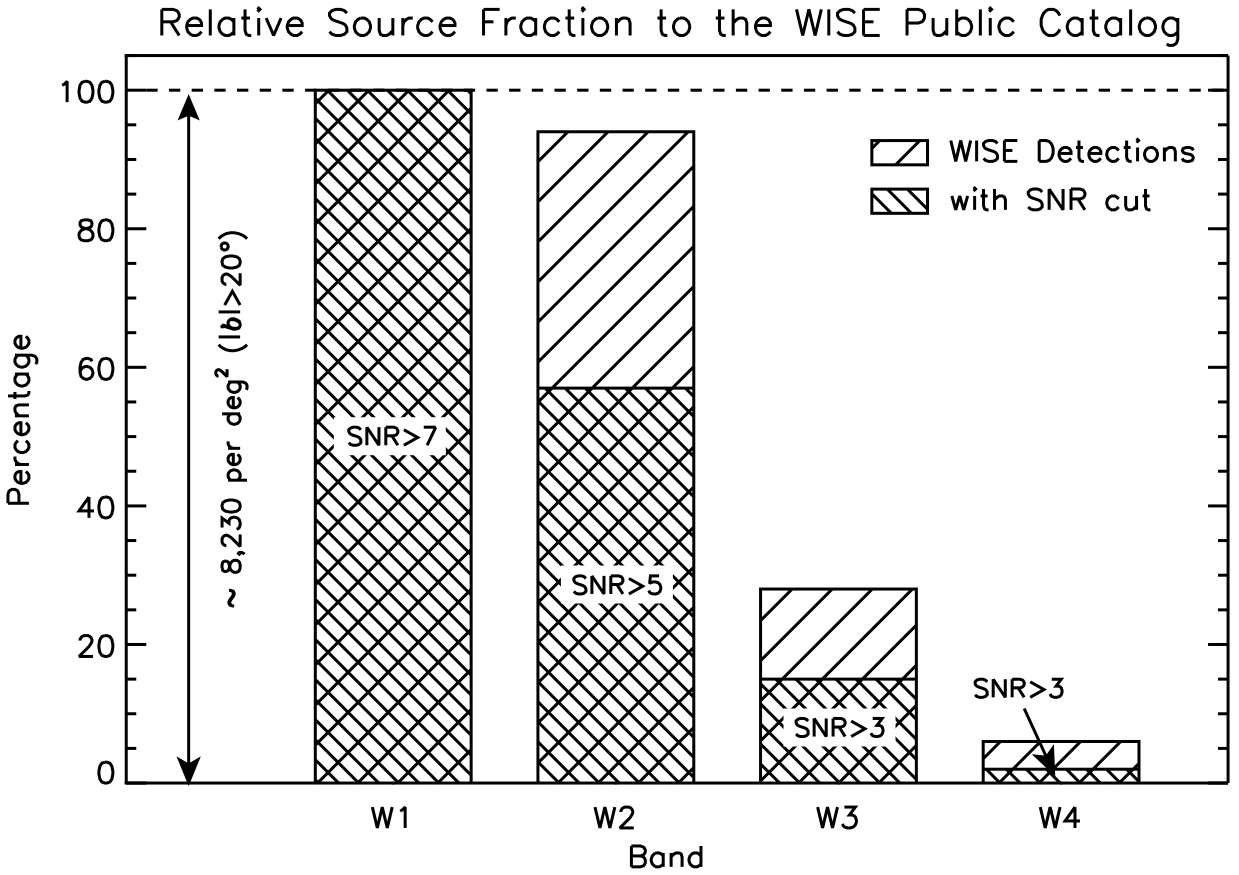}
%\plotone{bar0426.jpg}
\caption{Source densities of our SNR selected \wise\, samples for the four bands at $|b|>20^\deg$.  The left-most
bar, essentially equivalent to the 3.4\um\ SNR\,$\ge$7 catalog, has
a source density of 8230 per deg$^2$.  The
bars illustrate the source fractions with the corresponding SNR cuts at the four bands, relative to the original \wise\, preliminary public release source catalog, which includes all sources with
SNR $\geq$7 in at least one band. 
\label{bars}}
\end{figure}

Figure~\ref{bars} shows the source fraction in each of the above
three samples relative to the four-band merged, public released
catalog. When applying additional photometric SNR cuts, the source
sample sizes become increasingly smaller. At SNR\,$\ge$\,5, 4.6\um\
sources are only 57\%\ of the W1 sample.  At SNR\,$\ge$\,3,
this fraction is only 15\%\ and 2\%\ at 12\um\ and 22\um, respectively.
In the sense of source surface density, the W1 sample has
$\sim$\,8230 sources per deg$^2$, whereas the W1/2  and W1/2/3 
samples have only 4700 and 1235 sources per deg$^2$, respectively.
For comparison, the source surface density in the SDSS photometric
catalog is about 27,300 sources per deg$^2$, over three times higher
than that of \wise\, 3.4\um. This is because SDSS detects many more
low-luminosity (low-mass) galaxies than \wise\, \citep[see,
\eg][]{emilio12}.

For the three samples specifically selected for this paper,  SNR$_{W1}$\,=\,7 corresponds
to apparent magnitudes between 16.8 and 17.6\,mag (0.058\,-\,0.028\,mJy), SNR$_{W2}$\,=\,5 to 15.9\,mag to 16.5\,mag (0.075\,-\,0.043\,mJy), and SNR$_{W3}$\,=\,3 to the limiting magnitudes of 12.2 to 12.8\,mag (0.38\,-\,0.22\,mJy). 
%of 5 at 4.6\um\ corresponds to apparent magnitudes between 15.9 and
%16.5\,mag ($0.075 - 0.043$\,mJy).  Similarly, the $W3$ SNR3 sample
%has 12\um\ limits ranging from 12.2 to 12.8 mag ($0.24 - 0.42$\,mJy). 

\wise\, photometry is calibrated relative to measurements of standard stars, using the Vega magnitude system.  The conversion
factors to the AB system ($m_{\rm AB} \equiv m_{\rm Vega} + \Delta
m$) are 2.683 ($W1$), 3.319 ($W2$), 5.242 ($W3$) and 6.604 ($W4$);
equivalently, zero magnitude corresponds to 309.5, 171.79, 31.676
and 8.36\,Jy for the four bands, respectively \citep{ned2010} and \citep{Jarrett11}.  SDSS
photometry directly output from the SDSS archive is in the asinh
magnitude system.  We translate asinh magnitudes into the AB system
using information provided by SDSS\footnote{See {\it
http://www.sdss.org/DR7/algorithms/fluxcal.html.}}. When calculating
SDSS$-$\wise\, colors, we convert SDSS photometry to the Vega system
using $r{\rm (AB)} - r{\rm (Vega)} = 0.16$ \citep{Fukugita95}.
Throughout this paper, \wise\, magnitudes are in the Vega system,
SDSS magnitudes are in the AB system, but all colors are in the
Vega system.

%%%%%%%%%%%%%%%%%%%%%%

% FIGURE 2:  WISE source densities

\subsection{SDSS data\label{sec-sdssdata}}

This paper uses the SDSS DR7 data \citep{abazajian}, including
the photometric catalog, the main galaxy spectroscopic catalog\footnote{See
{\it http://www.mpa-garching.mpg.de/SDSS/DR7/.}} \citep{brinchmann04},
the luminous red galaxy (LRG) spectroscopic sample \citep{eisenstein01}
and the QSO sample \citep{schneider10}.  The DR7 legacy survey
catalog covers 8423 deg$^2$ and contains 230 million sources.  The main galaxy spectroscopic catalog 
is generated jointly by the Max-Planck-Institut fur  Astrophysik and the Johns Hopkins University (MPA-JHU DR7 main galaxy catalog; Brinchmann et al. 2004). It consists of almost $10^6$ galaxies
with \citet{Petrosian1976} magnitudes brighter than  $r = 17.77$ for
which various derived physical parameters are readily available.
We adopt the following classification criteria as in Kauffman et al. (2003): star-forming (SF)
galaxies are defined as having 
$\log[{\rm O~III}/{\rm H}\beta] < $1.3 + 0.61
$\log[{\rm N~II}/{\rm H}\alpha - 0.05]$
\citep[Eq.~1 in][]{kauffmann03}; AGNs (Seyfert and LINERs) are
defined as having $\log([{\rm O~III}]/{\rm H}\beta) > 1.19 + 0.61/[\log([{\rm N~II}]/{\rm H}\alpha) - 0.47]$ 
\citep[Eq.~5 in][]{kewley01}; and composite systems are defined as having
$\log([{\rm O~III}]/{\rm H}\beta)$ between the two values described
by the above equations.  These SF galaxy and AGN definitions are used in \S~\ref{sec-colors} when we utilize SDSS spectra
to classify \wise\ sources. When based on \wise\ data alone, we do not use these definitions. 
Two LRG samples are selected in color-magnitude
space, one to $r = 19.2$ (roughly volume-limited, to $z = 0.38$)
and one to $r = 19.5$ (flux-limited, to $z = 0.55$) \citep{eisenstein01}.
The QSO sample is selected by their non-stellar colors or FIRST
\citep{Becker1995} radio emission to $i = 19.1$ for $z < 3$ and
to $i = 20.2$ for $3 < z < 5.5$ \citep{richards02}.

In addition, all SDSS photometric redshifts (when spectroscopic redshifts are not available) used in sections below are the ones derived based on the neural network method \citep{oyaizu08}. We caution that, in general, broad band photometric redshifts are highly uncertain for strong type-1 AGNs/QSOs \citep{assef10,brodwin06}. These strong type-1 AGNs from SDSS have the most complete spectroscopic redshifts.

\section{Analysis and Results \label{sec-results}}

\subsection{Apparent magnitude and redshift distributions \label{sec-31}}

The \wise\, 3.4 and 4.6\um\ bands primarily sample emission from
stellar photospheres, whereas the 12 and 22\um\ bands are more
sensitive to dust emission heated by stars and accreting black
holes. Therefore, \wise\, readily identifies a variety of galaxy
populations in the mid-infrared sky.  To characterize these
populations, we first address several basic properties of the matched
\wise-SDSS catalog, including optical brightness distributions,
redshifts, the fraction of \wise\, sources undetected or with very
faint optical counterparts, and optical thru mid-infrared colors
of matched sources.

\begin{figure}[!h]
\epsscale{0.8}
\plotone{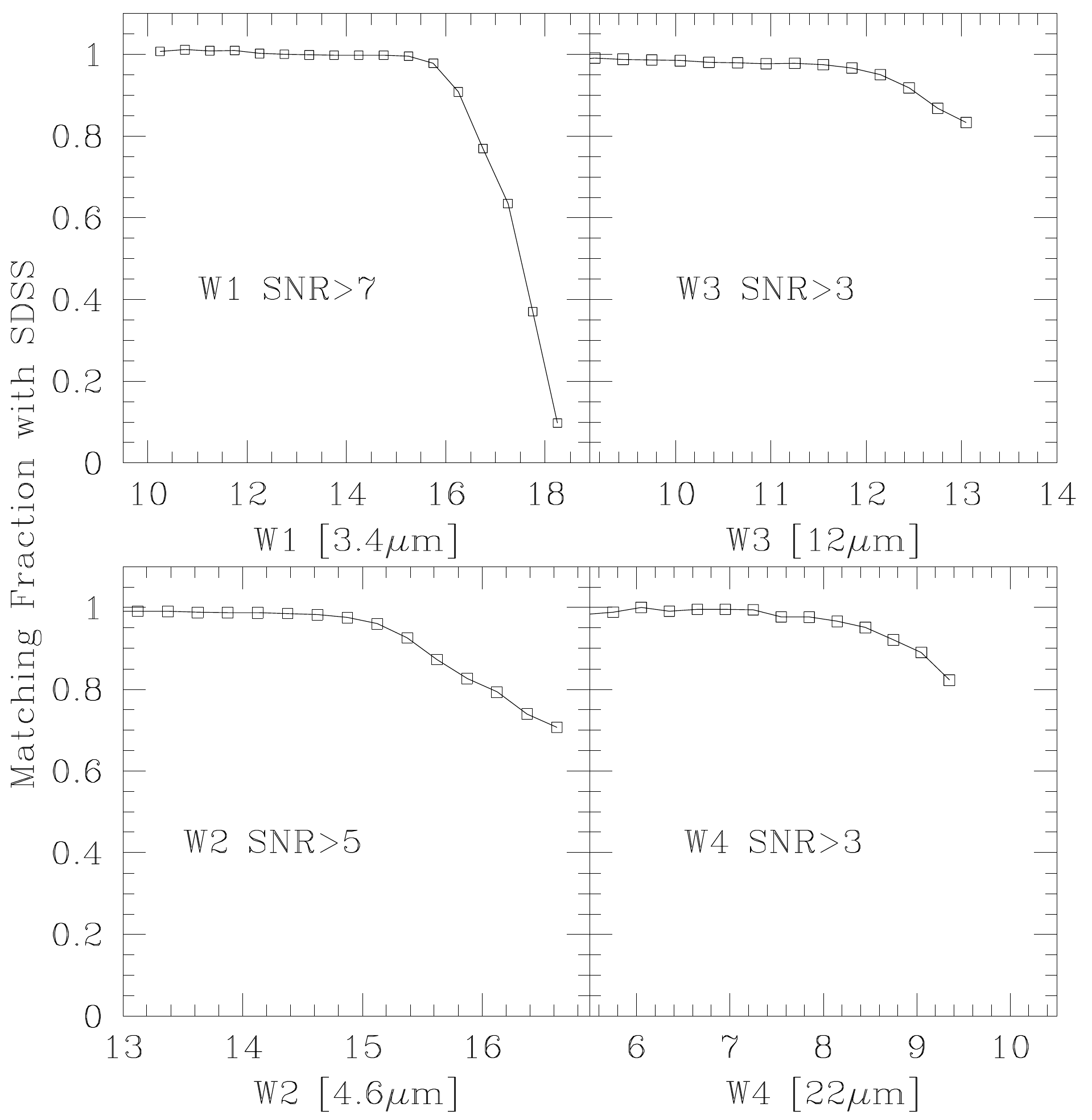}
\caption{Percentage of \wise\, sources matched with SDSS photometric
catalog as a function of \wise\, magnitude.
\label{match}}
\end{figure}

We carried out source matching between \wise\, and SDSS DR7 with a
matching radius of 3\arcsec.  This is based on the fact that the
\wise\, angular resolutions are 6\farcs1, 6\farcs4, 6\farcs5, and
12\arcsec\, for the four bands, respectively.  Figure~\ref{match}
shows the percentage of \wise\, sources with SDSS $r$-band optical
counterparts for the four \wise\, bands.  To a depth of $r = 22.6$,
corresponding to the 50\%\ completeness depth of SDSS \citep{abazajian04},
91\%\ of the W1/2/3 and 96\%\ of the W1/2/3/4 (SNR$_{W4}\ge3$)
samples have optical counterparts.  In contrast, only 72\%\ and
86\%\ of the W1  and the W1/2 samples have optical counterparts
with $r < 22.6$. This implies that the bulk of 12 and 22\um\ galaxies
are dusty star-forming galaxies and AGNs at low redshifts ($z \sim
0.1 - 0.3$), whereas 15\,-\,25\%\ of 3.4 and 4.6\um\ sources could
be massive early-type galaxies which are fainter than the SDSS
photometric limits and are at $z$\,$\ge$\,1.  Indeed, \citet{gettings12}
reports on a \wise-selected galaxy cluster at $z=0.99$ which is
well-detected in $W1$ and $W2$ but is undetected by SDSS.

\begin{figure}[!h]
\plotone{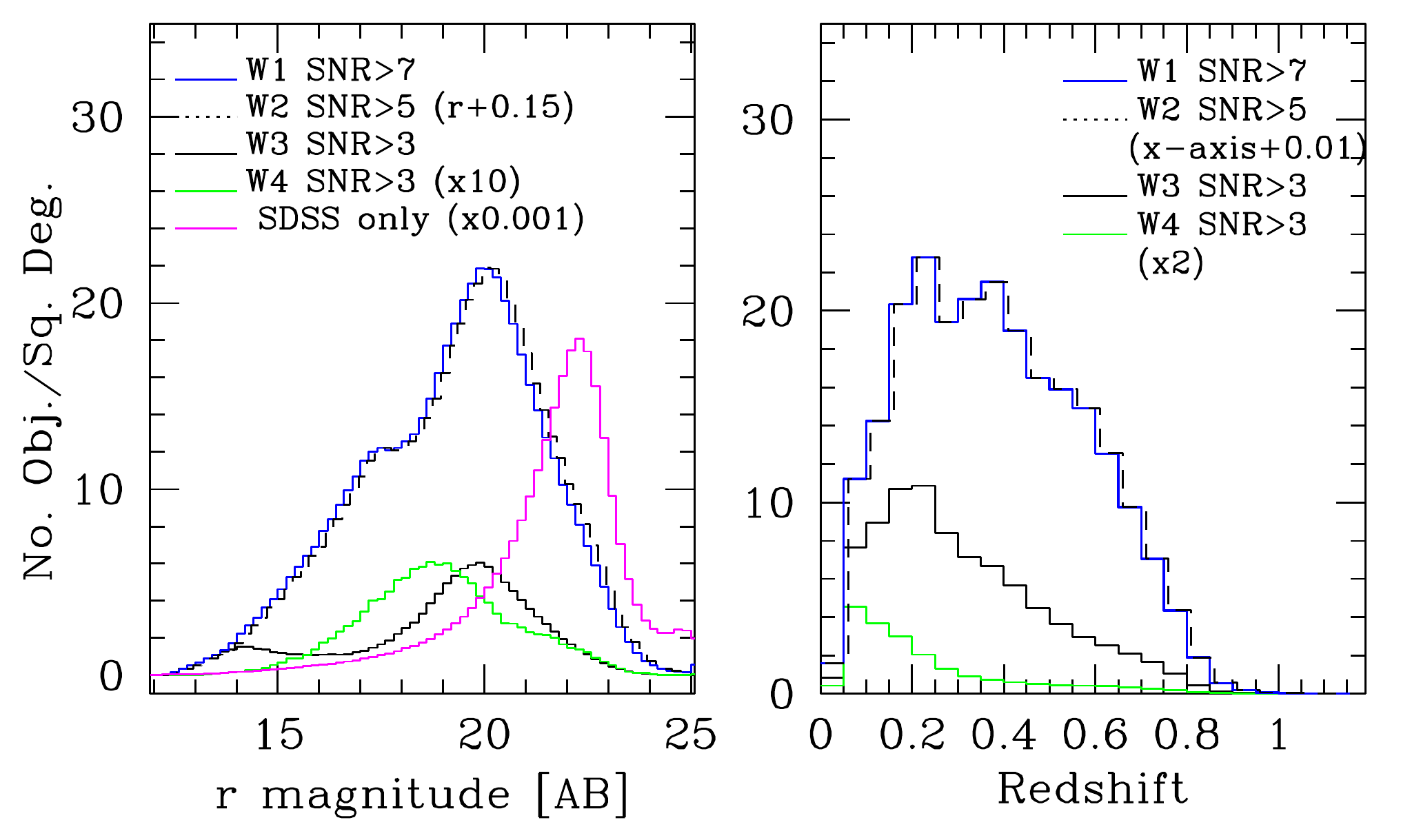}
\caption{The $r$-band magnitude and photometric redshift distributions
for the WISE-SDSS matched sources from the four \wise\ samples.  For both panels, 
the distributions for W1 and W1/2 samples are {\it indistinguishable}, with the W1 sample 
in blue and the W1/2 sample is green. For the visual clarity, we slightly shifted
the W1/2 sample along the x-axis as labeled in the figure legend.
In the left panel, 
for the W1/2/3/4 sample, 
we scaled the y-axis by a factor 10 for visual clarity.  Similarly, in the right panel, the
y-axis for the W1/2/3/4 sample is scaled by 2.
 For comparison, we also plot the $r$-band magnitude
distribution for the full SDSS photometric catalog (scaled down by
a factor of 1000).  Note that SDSS optical magnitudes are in the
AB system, calculated from SDSS asinh magnitudes.  The $r$-band
magnitude histograms have a bin size of 0.2\,mag.  The majority of
12\um\ sources have bright SDSS optical counterparts and are at low
redshifts ($z \sim 0.2$).
\label{rzhistogram}}
\end{figure}

Figure~\ref{rzhistogram} shows the optical brightness and photometric
redshift distributions for the mid-infrared sources detected in the four bands. 
%Here we apply criteria of SNR$>$7 for $W1$, SNR$>$5 for $W2$, SNR$>$3 for $W3$ and $W4$.
It is clear that \wise\ 22\um\ sources have corresponding optical magnitudes which 
peak at $r\sim$19, one magnitude brighter than the peaks for the W1, W2 and W3 samples. 
For comparison, the parent SDSS
photometric catalog continues to rise until completeness causes an
apparent drop at $r \sim 22$.  This confirms that SDSS detects many
more low-mass, low-luminosity galaxies than \wise\, and the bulk
of \wise\, galaxies with optical counterparts are relatively local,
at $0.1 < z < 0.3$.

\begin{figure}[!h]
\plotone{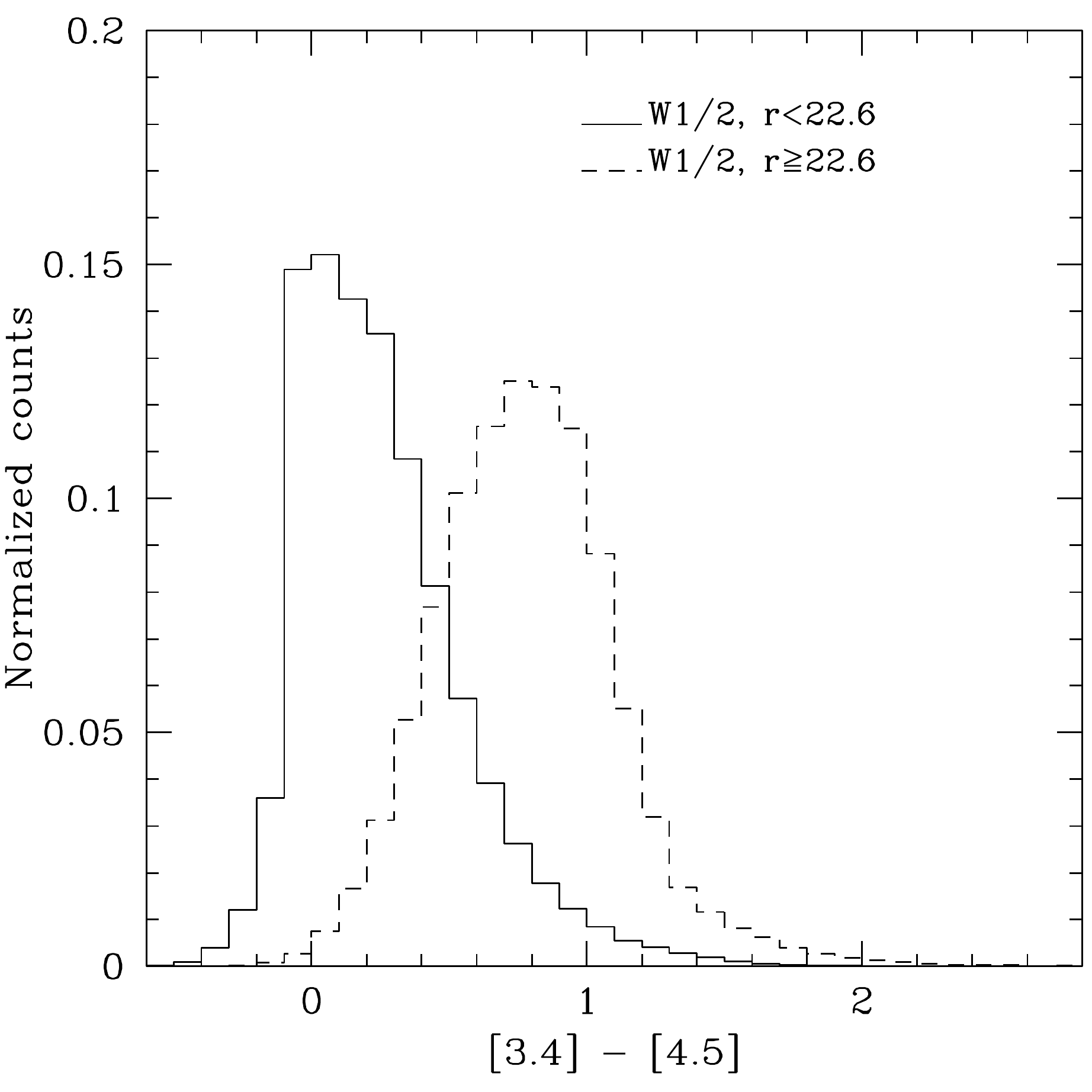}
\caption{Normalized $W1 - W2$ color distribution for W1/2 sample 
with bright optical counterparts (solid line; $r < 22.6$) compared
to sources with faint or undetected optical counterparts
in the SDSS data ($r \geq 22.6$).
\label{w12histogram}}
\end{figure}

About 28\%\ and 14\%\ of \wise\, 3.4\um\ and 4.6\um\ sources are
very faint or without any optical counterparts in the SDSS photometric
catalog.  These optically faint 3.4\um\ galaxies are predominantly
massive early-type galaxies at $z \simgt 1$, though heavily obscured
galaxies and AGN also contribute at some level.  Figure~\ref{w12histogram}
examines the $W1 - W2$ colors of this optically faint population,
showing that $W1$ sources with very faint $r$-band magnitudes ($r
> 22.6$) have redder $W1 - W2$ colors than optically brighter $W1$
sources.  As discussed below and shown with color-redshift tracks
for galaxy templates from \citet{assef10}, redder $W1 - W2$ color
is an indicator of early-type and star-forming galaxies being at
higher redshifts, $z \simgt 1$.

% FIGURE 3:  WISE-SDSS matching

% FIGURE 4:  r-band mag distributions

% FIGURE 5: W1-W2 color distributions

\begin{figure}[!h]
\plotone{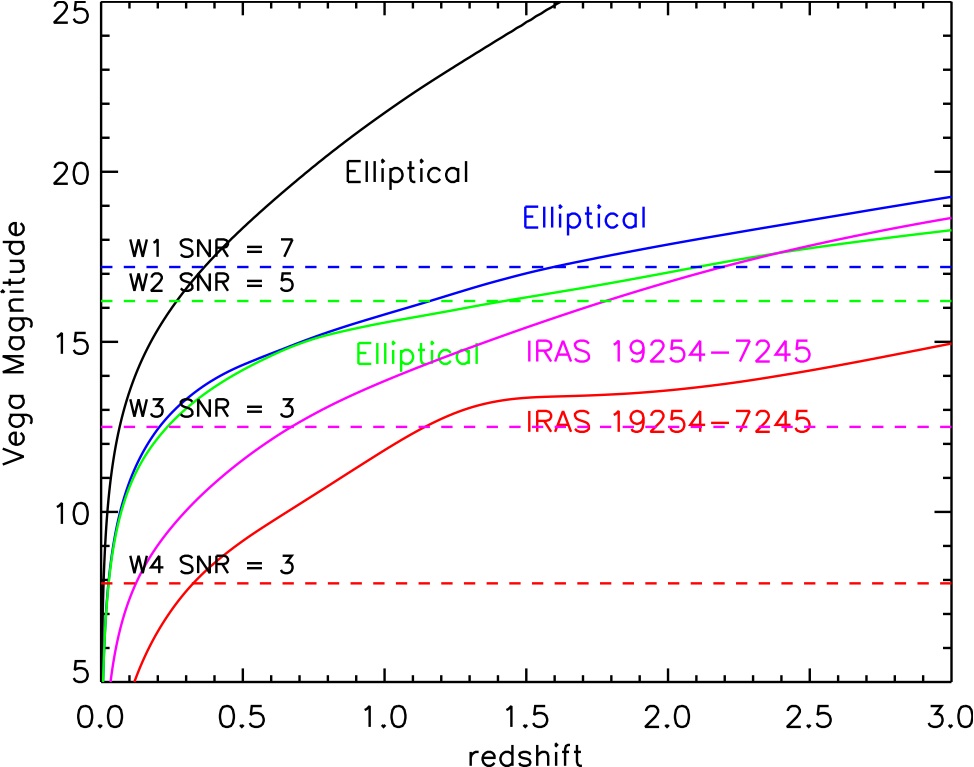}
\caption{Predicted $r$-, $W1$-, $W2$-, $W3$- and $W4$-band magnitude
as a function of redshift (corresponding to black, blue, green, magenta and red lines respectively) using two empirical  galaxy SED templates, elliptical galaxy and IRAS19254-7245. 
For the early type galaxy, the template is normalized to an L$^*$ galaxy at $z=0.043$ with $\rm L_K=3.5\times10^{11}L_\odot$ \citep{Lin04,mancone10}.  Minimum \wise\,
sensitivity limits are shown as dashed horizontal lines.  This shows
that \wise\, 3.4\um\ images are sensitive to typical early-type
galaxies out to $z \sim 1$; this sensitivity is enhanced for more
luminous early-type galaxies and/or higher ecliptic latitude fields
where the \wise\, coverage is deeper.
\label{magz}}
\end{figure}

What types of galaxies are detected by \wise\, at 3.4 and 22 microns?
One way to answer this is to calculate expected \wise\, magnitudes
as a function of redshift using empirical galaxy templates.
Figure~\ref{magz} shows the expected $W1$ and $W2$ magnitudes for
an early-type galaxy template, as well as $W3$ and $W4$ magnitudes
for an infrared-bright galaxy (IRAS~19254$-$7245) as a function of
redshift.  
%\wise\, magnitude limits are marked with dashed horizontal
%lines.  The early-type galaxy model is normalized to an L$^*$ galaxy with rest-frame
%$\rm L_K=3.5\times10^{11}\, L_\odot$ at $z=0.043$ \citep{Lin04, mancone10}. 
This figure suggests that ignoring evolution, L$^*$ early-type galaxies should be visible in \wise\ 3.4\um\ data
out to redshifts of 1.5\,-\,2.  Another important result emphasized by
Figure~\ref{magz} is that the observed $W1$ magnitudes do not change
significantly over $z \sim 0.5 - 1.5$ for early-type galaxies; this
benevolent $k$-correction has been noted repeatedly for the similar
3.6\um\ band of \spitzer\ \citep[\eg][]{eisenhardt08, mancone10,
galametz12}.  In contrast, optical $r$-band magnitudes steeply
decline in brightness with redshift.

\begin{figure}[!h]
\plotone{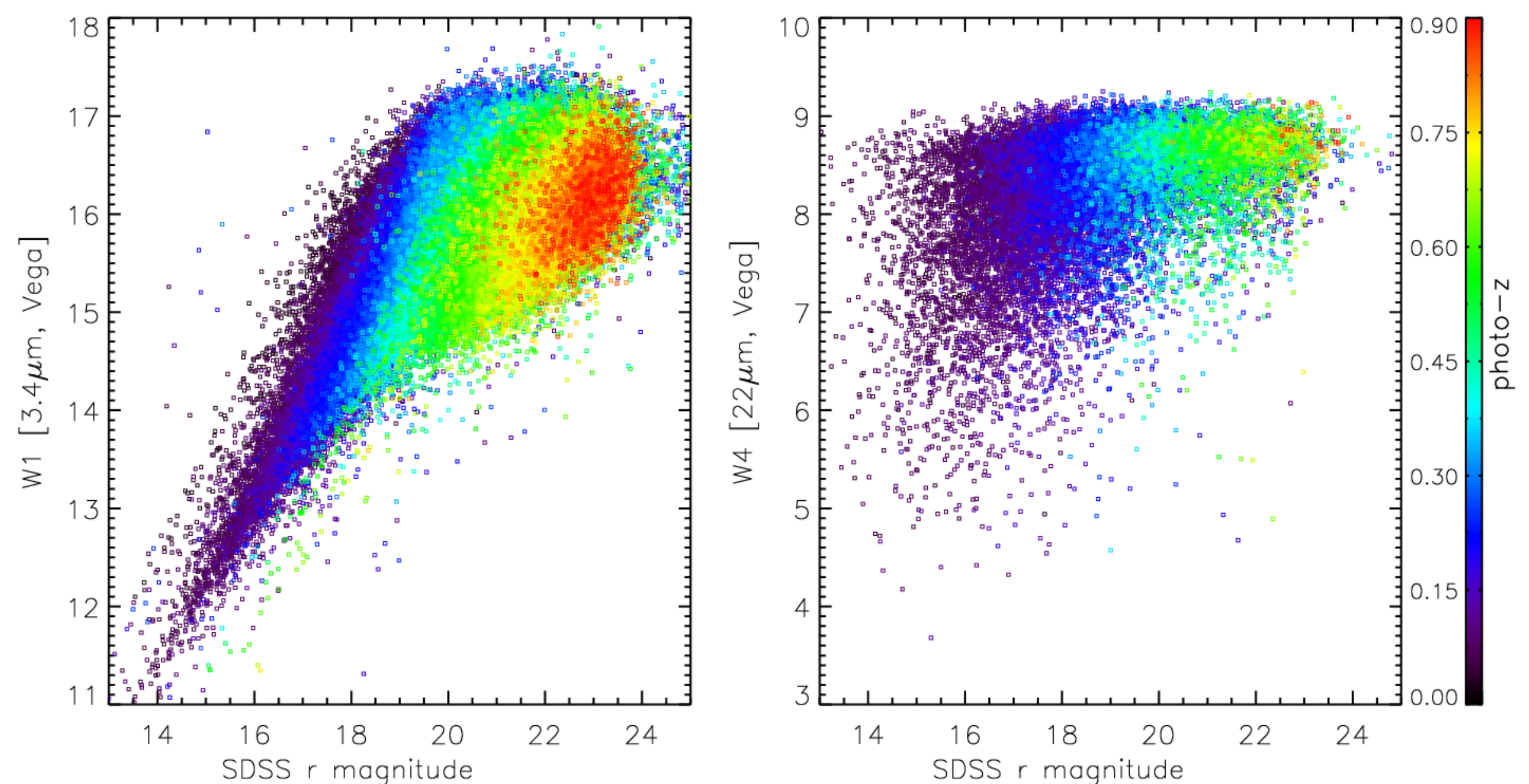}
\caption{\wise\, 3.4\um\ and 22\um\ magnitude as a function of
optical $r$-band magnitude. The colors indicate photometric redshifts
taken from the SDSS catalog. As expected, optically fainter galaxies,
\ie\ sources with redder $r - W1$ or $r - W4$ colors tend to be at
higher redshifts.
\label{faintr1}}
\end{figure}

Figure~\ref{faintr1} illustrates how $W1$ and $W4$ magnitudes change
with optical brightness, $r$, and photometric redshift.  $W1$ traces
$r$ until $r \sim 20 - 21$, at which point $W1$ magnitudes remain
within a narrow range of 15.8\,-\,16.8\,mag, while optical brightness
becomes increasingly fainter, stretching over three magnitudes.
Figure~\ref{magz} and Figure~\ref{faintr1} illustrate the same
effect, \ie\, most galaxy SEDs have very steep slopes at ultraviolet
through optical wavelengths, and much shallower slopes in the
near-infrared.  The turn-over at $r \sim 20$ in Figure~\ref{faintr1}
implies that the bulk of $W1$ sources are low-redshift ($z < 0.5$),
typical SDSS-detected galaxies, and $W1$ sources with $r \simgt 20$
tend to be at higher redshifts ($z \sim 0.5 - 2$).

% FIGURE 6:  model diagram

% FIGURE 7:  mag-mag diagram
\epsscale{1.1}
\epsscale{1.0}

\subsection{Separating powerful AGN/QSOs at $z\le3$ from galaxies \label{sec-colors}}

\begin{figure}[!htr]
\plotone{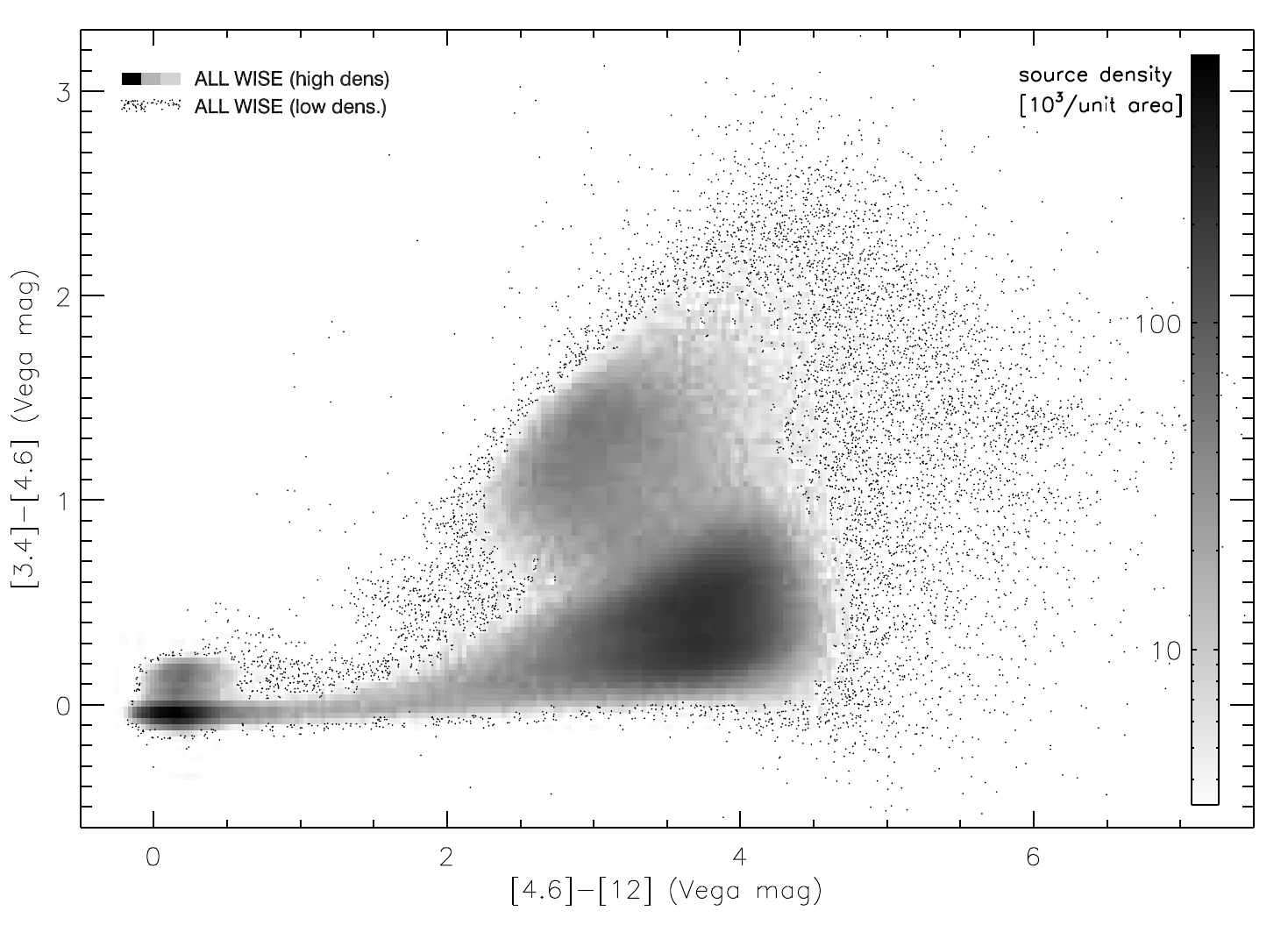}
\caption{Color-color distribution of \wise\, sources from the
12\um-selected W1/2/3 sample.  To illustrate the large range in
source density in color-color space, we combine a grey-scale plot
for high concentration regions (see scale-bar on right) and individual
points for low source density regions.
\label{wcc}}
\end{figure}

With four-band mid-infrared photometry, can \wise\, colors alone
provide diagnostics for different types of sources as well as crude
redshift information?  Figure~\ref{wcc} addresses this question by
showing $W1 - W2$ vs. $W2 - W3$ colors for the W1/2/3 sample.  The three
high source density regions are:  {\it (1)} the stellar locus with
both colors near zero; at mid-infrared wavelengths, emission from
Galactic stars from spectral class O through $\sim$~T3 is dominated
by the Rayleigh-Jeans tail of the blackbody spectrum, thereby
yielding Vega-system colors near zero for most Galactic stars (note
that pure elliptical galaxies with no dust at $z \leq 0.1$ also sit
near the stellar locus); {\it (2)} the red $W1 - W2$ cloud with
$W1 - W2 > 0.8$; and {\it (3)} the bluer $W1 - W2$ sequence, spanning
a wide range in $W2 - W3$ color.   Here we do not see any extremely
cool brown dwarfs with very red $W1 - W2$ color due to methane
absorption since such sources are very faint in $W3$, and thus are
excluded in our W1/2/3 sample \citep{davy11}.

\begin{figure}[!htr]
\plotone{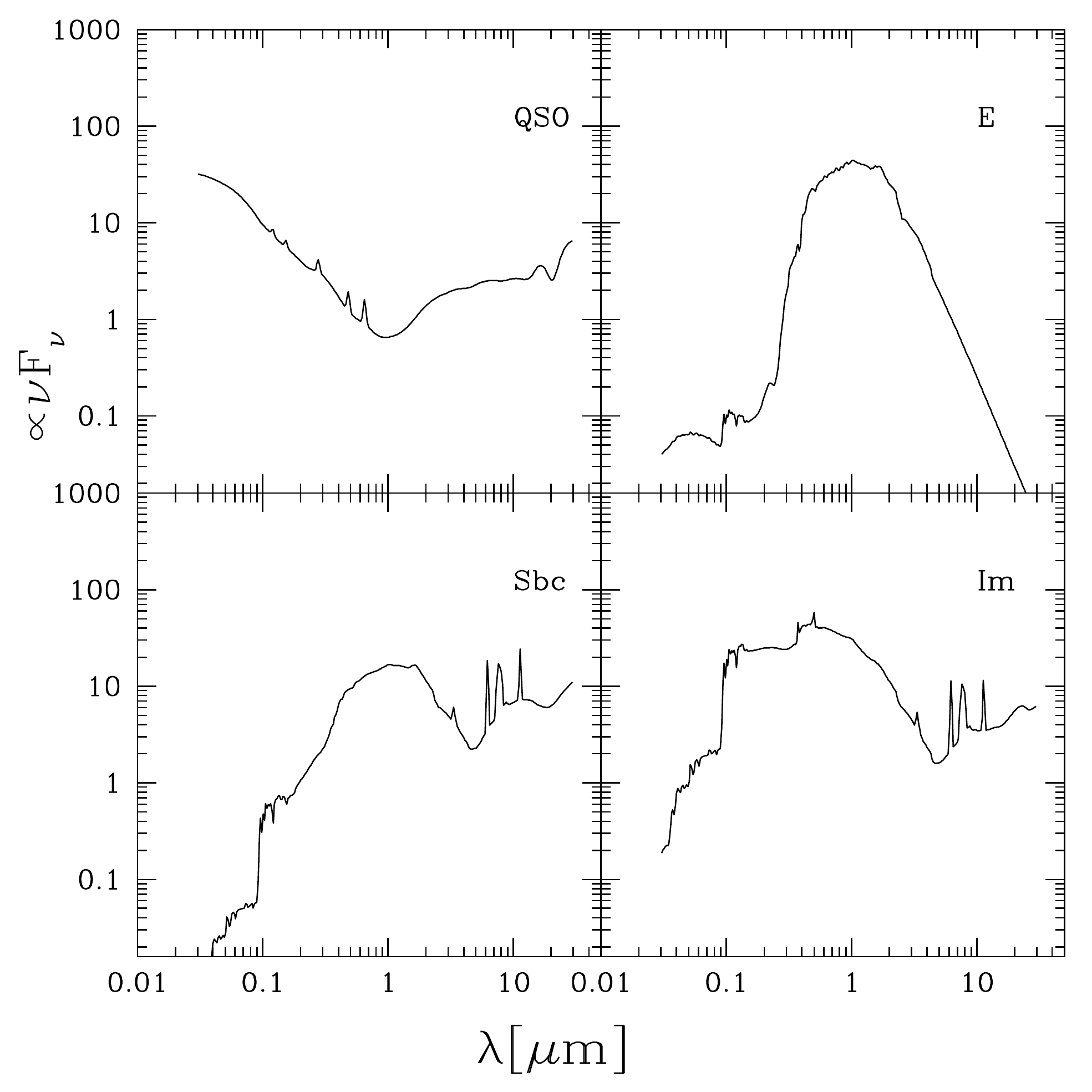}
\caption{Empirical templates for quasars and normal galaxies from
\citet{assef10}.  The major difference between galaxies and AGN is
at 1-2 \um:  AGN have a minimum in that wavelength range, while
galaxies have a peak in that wavelength range.  Star-forming and
passive galaxies are easily distinguished from their longer-wavelength
data ($\simgt$ 4\um).
\label{template}}
\end{figure}

\begin{figure}[!htr]
%\plotone{/Users/linyan1/wise-science/papers/data/plots/SEDtracks0426.jpg}
\plotone{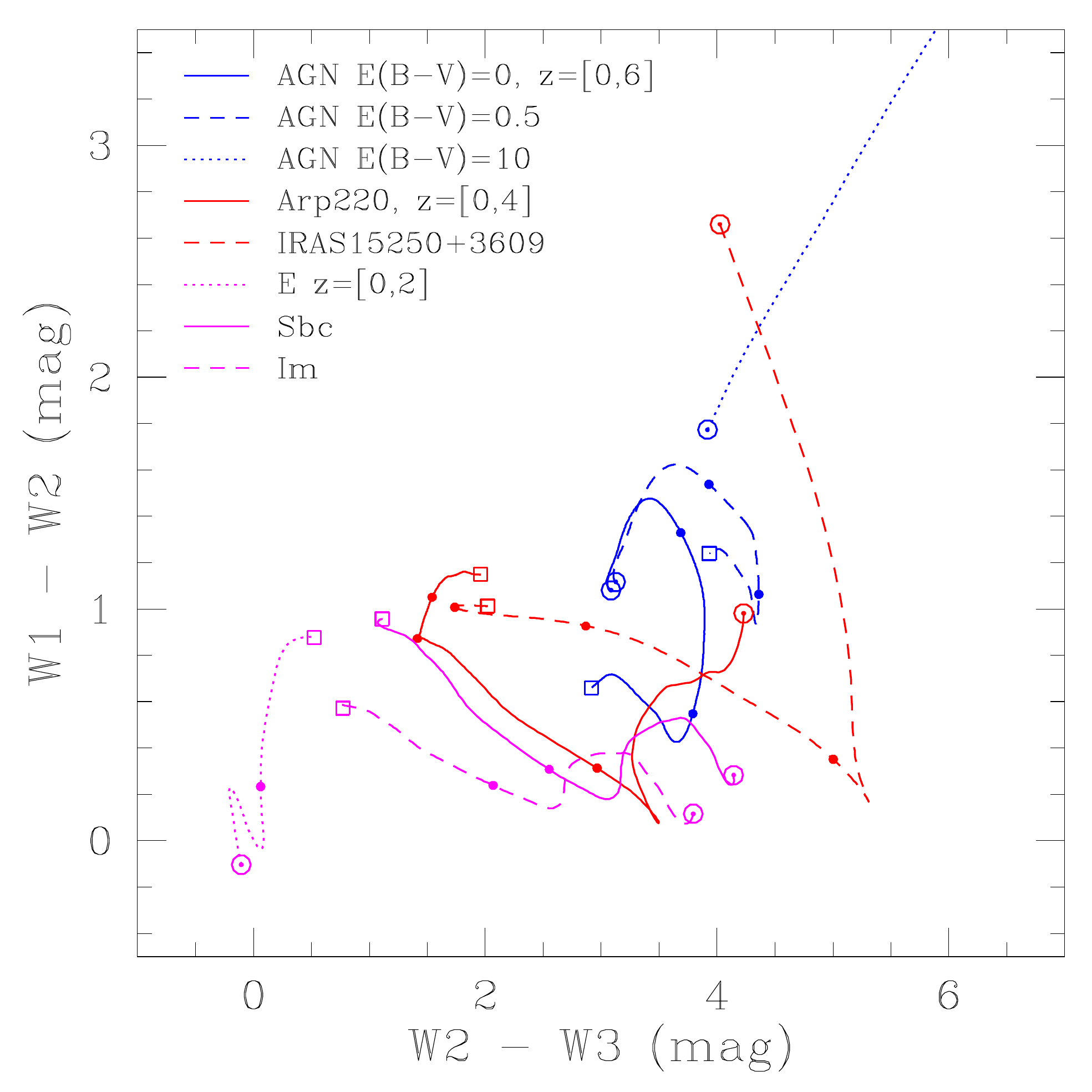}
\caption{Color-color tracks as a function of redshift for several
galaxy and AGN templates taken from Polletta et al. (2007), Vega et al. (2008) and Assef et al. (2010). The blue tracks are for AGNs with target symbols indicating $z=0$, black dots for $z=2,4$ and open squares for $z=6$.  The AGN tracks are shown with three different dust obscuration factors. The red tracks are for the local ULIRGs Arp\,220 and IRAS15250+3609 with target symbols for $z=0$, solid dots for $z=1,2,3$ and open squares for $z=4$.  The magenta tracks are for normal galaxy templates with target symbols for $z=0$, solid dots for $z=1$ and open squares for $z=2$. \label{modelcc}}
\end{figure}

\begin{figure}[!h]
\plotone{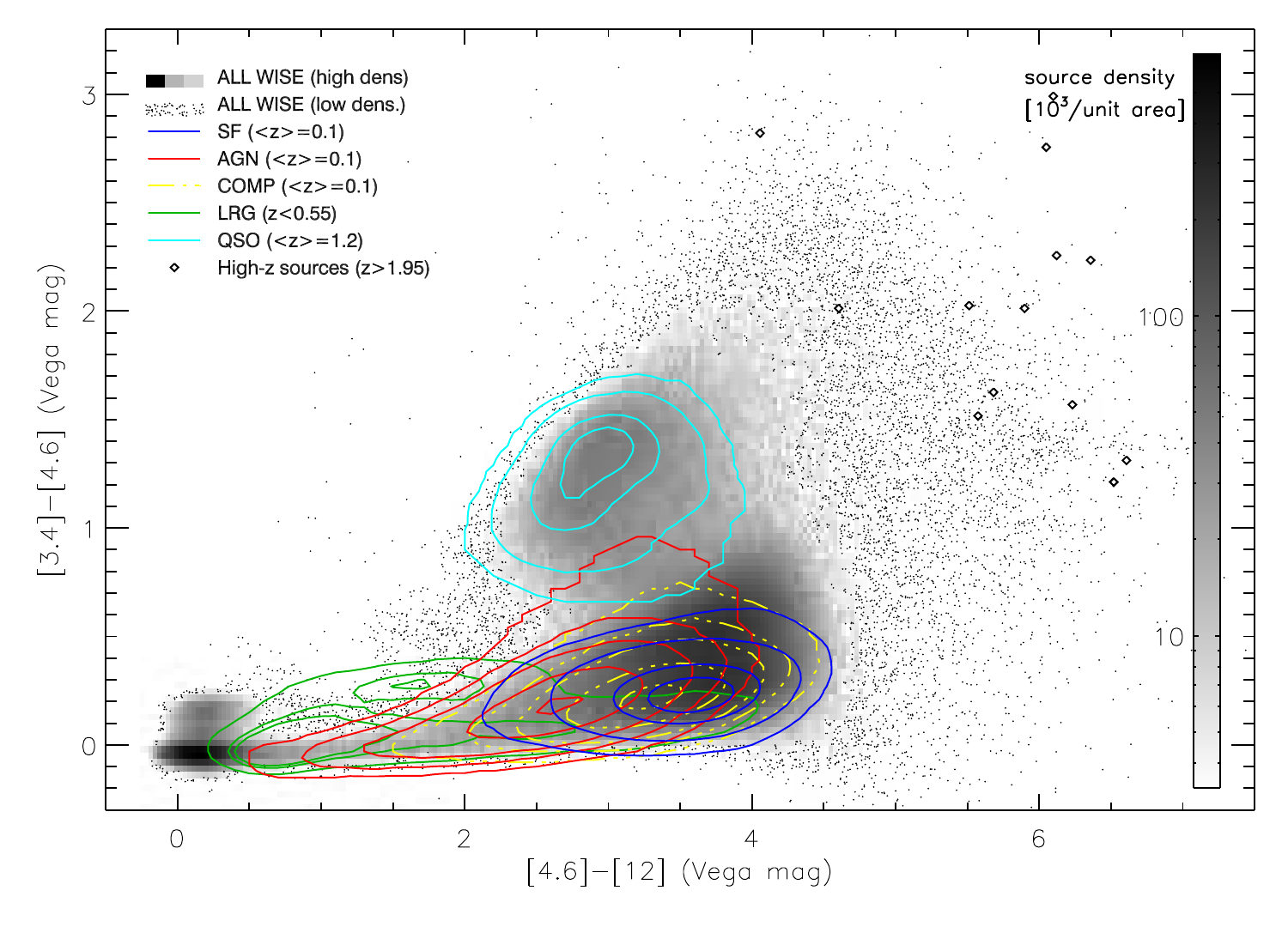}
\caption{As per Figure~\ref{wcc}, with source classifications
indicated, based on the SDSS DR7 spectroscopic galaxy catalog
(see text for details). Spectroscopic confirmed QSOs and LRGs are also
plotted. The color contours indicate sources which have been
classified as star forming galaxies (SF, blue), seyfert AGNs (red), composite systems (yellow), LRGs (green), 
and bright QSOs (cyan).  These color contours are used to visually illustrate the concentrations of source distributions in this color-color space. Additionally, we plot several extremely red \wise-selected
ULIRGs at $z \simgt 2$ from \citet{eisenhardt12}, Stern et al. (in prep), \citet{bridge12}, \citet{jingwen12},
and \citet{yan12}.
\label{wcc2}}
\end{figure}

What is the physical basis of the color-color distribution shown in Figure~\ref{wcc}? Figure~\ref{template}
through \ref{wcc2} examine this question from different angles.
Figure~\ref{template} shows empirical galaxy and AGN templates from $0.03-30$\um\ based on $\sim20000$ objects with multi-wavelength data in NOAO Deep Wide-Field Survey \citep{assef10}.  For systems with strong nuclear heating, the
mid-infrared SEDs tend to be roughly a rising power-law, resulting
in $W1 - W2 > 0.8$.  For systems with dominant stellar emission at
$z < 1$, this color is smaller because the 3.4\um\ band is sampling
below the 1.6\um\, H$^-$ opacity peak of the stellar emission, making the $W1 -
W2$ color bluer; at $z > 1$, the 3.4\um\ band samples below the
near-infrared stellar peak, getting fainter, thereby providing
redder $W1 - W2$ colors.

Figure~\ref{modelcc} shows the $W1 - W2$ and $W2 - W3$ color tracks as a function of redshift calculated using the set of SED templates of Assef et al. (2010), the Arp\,220 SED template of \citet{polletta07} and the IRAS15250+3609 template of \citet{vega08}.  The model
colors clearly do not span the full range of observed values in
Figure~\ref{wcc}.  The assumed SED templates, generated from
optical and \spitzer\, observations, are by no means complete. These templates include elliptical galaxies, 
star forming Sbc and Im type galaxies,  type-1 AGNs and local ULIRGs Arp\,220 and IRAS15250+3609.
The AGN tracks have three different dust obscuration factors; dust-obscured ones are type-2 AGNs.
%We only plot tracks down to the limit of
%the W1/2/3 sample.  

%Comparing Figure~\ref{modelcc} and \ref{wcc2},
%we see that the newly discovered \wise-selected $z$\,$\sim$\,2
%ULIRGs have much redder $W2 - W3$ colors than the calculated high-$z$ colors using local ULIRG SED templates, including
%the reddest one such as IRAS15250+3609.  This point
%is thoroughly addressed in several companion papers by the \wise\,
%extragalactic team using full infrared SEDs of such sources from
%ground-based sub-millimeter and/or \herschel\, five-band far-infrared
%photometry \citep[\eg][]{bridge12, eisenhardt12, stern12b, jingwen12,
%yan12}.

As opposed to Figure~\ref{modelcc} which shows \wise\ colors 
based on SED templates, Figure~\ref{wcc2} is the color-color plot for
known sources. Specifically, we plot SDSS spectroscopically classified
sources, including star-forming (SF) galaxies, galaxies hosting
AGN, SF/AGN composite systems, LRGs, QSOs and $z \simgt 2$
\wise-selected ULIRGs.  These figures enable us to conclude that
\wise\, colors alone are effective in separating strong AGNs from
star-forming galaxies, and in separating stars from extragalactic
sources (other than early-type galaxies at very low-redshift).

\begin{figure}[!h]
\plotone{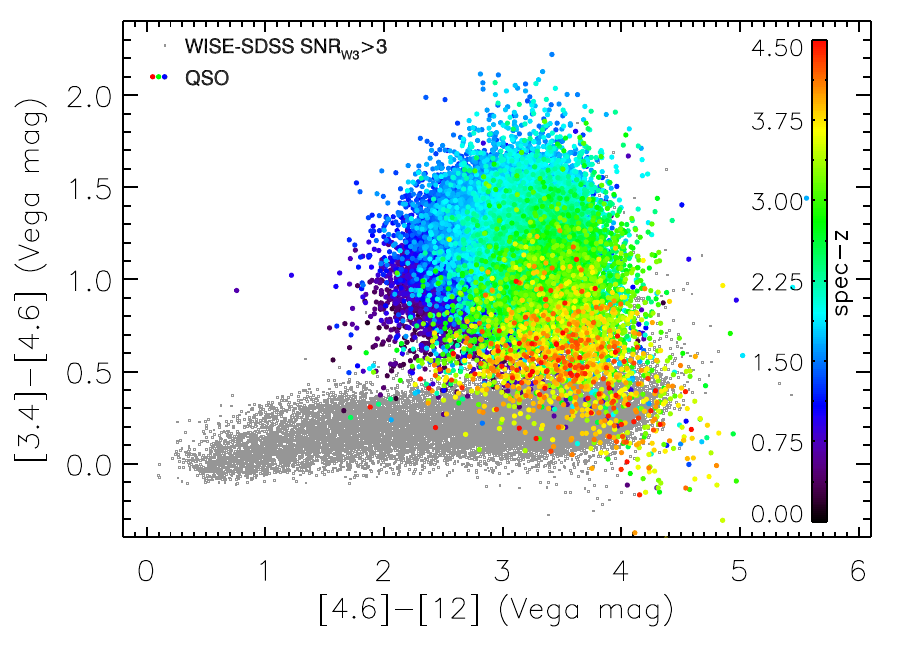}
\caption{\wise\, color-color distribution for SDSS spectroscopically
confirmed quasars, color-coded by redshift. Grey points are normal
star-forming galaxies.  A simple $W1 - W2$ color cut efficiently
and robustly separates quasars at $z \simlt 3$ from normal galaxies.
Higher redshift quasars, however, enter into the galaxy locus in
this color-color diagram.
\label{qso}}
\end{figure}

Figures~\ref{wcc} through \ref{wcc2} all suggest one unique application
of \wise\, data over the entire sky --- selecting strong AGN/QSO
candidates at $z < 3$ using $W1 - W2$ color.  Using mid-IR colors to select AGNs has been
noted  earlier in \spitzer\, studies over relatively
small areas and with much smaller AGN samples \citep[\eg][]{lacy04,
stern05}.
The selection of $W1-W2$ method, briefly discussed in \citet{Jarrett11}, is tested and
discussed in detail in \citet{stern12} and \citet{assef12} \citep[see
also][]{ashby09,assef10, eckart10, edelson12, massaro12, wu12,mateos12}.  
With the current analysis, we, for the first time,
investigate this simple selection criterion using the wide-area,
large spectroscopic database of SDSS.  $W1
- W2$ vs. $W2 - W3$ is plotted in Figure~\ref{qso} for SDSS spectroscopically confirmed QSOs,
color-coded by redshift. Normal, star-forming galaxies are also
plotted.  We see that QSOs at $z \le 3$ mostly have $W1 - W2 \ge
0.8$, and are reasonably well separated from star-forming galaxies.
However, $z \sim 3 - 5$ QSOs have bluer $W1 - W2$ colors due to the
3.4 and 4.6\um\ filters sampling rest-frame optical wavelengths,
making such quasars difficult to distinguish from normal galaxies based on mid-IR colors alone.
A similar issue with \spitzer\ selection of distant quasars was
pointed out by \citet{assef10}.  The selected AGN candidate sample
has two unique features: {\it (1)} it provides a more complete
selection of QSOs, particularly at $z \sim 2 - 3$, where the SDSS
optical color method has issues \citep{richards02,schneider10}; and
{\it (2)} it is less affected by dust than optical selection methods,
and thus is more sensitive to obscured, type-2 AGNs.

%%%% This paragraph needs revision%%%%%%%%%%%%
%Using mid-infrared colors to separate strong AGNs/QSOs from normal
%galaxies was earlier noted in \spitzer\, studies over relatively
%small areas and with much smaller AGN samples \citep[\eg][]{lacy04,
%stern05}.  
In addition to the color criterion of $W1-W2>0.8$, another important condition for selecting AGN with high completeness
and reliability is $W2$ magnitude and SNR limits (see \citet{stern12} and \citet{assef12} for the full details).
%The \wise\ color selection for AGN is discussed and tested in \citet{stern12} and \citet{assef12} in detail. 
Using the deep, multi-wavelength data in Cosmic Evolution Survey \citep[COSMOS;][]{scoville07}, Stern et al. (2012)
found that $W1-W2>0.8$ in combination with $W2<15.05$ at $SNR>10$ identifies 78\%\ of \spitzer\ mid-IR AGN candidates (as the truth sample) with a reliability of 95\%.
%The specific  $W1 - W2$ value of 0.8 was derived to separate AGN candidates from normal
%galaxies based on a balance between completeness and reliability using the deep, multi-wavelength data available in the Cosmic
%Evolution Survey \citep[COSMOS;][]{scoville07}.
%\citet{stern12} motivates how the specific $W1 - W2$
%value of 0.8 was derived to separate AGN candidates from normal
%galaxies, based on a balance between completeness and reliability
%and using the deep, multi-wavelength data available in the Cosmic
%Evolution Survey \citep[COSMOS;][]{scoville07}.  
%Adopting the
%\citet{stern05} mid-infrared AGN selection criteria as the truth
%sample, this simple $W1 - W2$ color cut identifies 78\%\ of \spitzer\
%mid-infrared AGN candidates with a reliability of 95\%\ down to the $W2$ depth of $<15.2$ at $>10\sigma$.  
%These
%numbers are determined from the \wise\ data in the COSMOS field with
%a depth of $10\sigma$ limit of $W2=15.2$ ($>$160$\mu$Jy). 
%a depth of 160 $\mu$Jy at 4.6\um\, ($W2 < 15.2$),
%corresponding to the $10\sigma$ depth of the \wise\, COSMOS data.
%The derived AGN candidate density is $61.9 \pm 5.4$ AGN candidates
%per deg$^2$, with a relatively large error bar due to small number
%statistics.  
\citet{assef12} uses higher latitude \wise\, data in
the wider-area Bo\"otes field to further investigate the completeness and reliability of the \wise\ AGN selection as functions of magnitude and SNR.  As shown in Figure~\ref{modelcc} and discussed in \S~\ref{sec-31}, optically faint, high-redshift elliptical and Sbc galaxies tend to have very red $W1 - W2$ colors. These sources are contaminants to \wise\ AGNs based on a pure $W1 - W2>0.8$ color cut.  To limit the contamination, Stern et al. (in prep) and Assef et al. (2012) found that  $W2<15.2$ at the number of sky coverage $\ge15$ can achieve completeness of 74\%\ and reliability of 89\%.  
Here similar to Stern et al. (2012), the reliability and completeness of \wise\ AGN sample is derived based on using the \spitzer\ color selected mid-IR AGNs as the truth sample. For our chosen \wise\ + SDSS overlap sub-region, more than 80\%\ of the sky have the coverage $\ge15$, and 99\%\ have the coverage $\ge10$. At $W2=15.2$, the median source SNR and the number sky coverage are 12 and 22 respectively. Therefore, for our analysis of \wise\ AGNs throughout this paper, we adopt $W2<15.2$ criterion. Our AGN selection criteria of $W1-W2>0.8$, $W2<15.2$ should maintain a reasonable completeness and a good reliability.

%Because \wise\ images the sky with different coverage, the magnitude limit of $W2<15.2$ does not correspond to the same SNR limit. At the lower SNR, there will be more contaminants in \wise\ color selected AGNs. Figure~\ref{xx} shows the reliability and completeness derived for the sample selected with $W1-W2>0.8$ and $W2<15.2$ as a function of the number of sky coverage. The x-axis is equivalent to $W2$ SNR.  In our analysis of \wise\ AGNs throughout this paper, we adopt $W2<15.2$ criterion. Within the \wise\ and SDSS overlapping area, this magnitude limit has SNR$>XX$. Thus our AGN selection criteria of $W1-W2>0.8$, $W2<15.2$ at SNR$>XX$ should maintain a completeness of xxx and a reliability of XXX. 

What is the source density of \wise-selected sources?  In the large,
high Galactic latitude \wise-SDSS overlap area we analyze here, the
W1/2/3 sample contains \,12\%\, Galactic stars, 12\%\ luminous
AGN candidates, and 70\%\ normal galaxies (possibly containing weak
AGNs).  The remaining 6\%\ of $W3$ sources are extremely red, with
either $W2 - W3 \ge 4.5$ or $W1 - W2 \ge 1.8$.  These present a
rare population of red objects, and are likely dusty systems at
high redshift (see \S~\ref{sec-highz}).  With the selection of $W1-W2>0.8$, $W2<15.2$, the \wise\ AGN/QSO candidates have a surface density of $67.5\pm0.14$ per deg$^2$.  Considering
only the optically bright subsample, with $r < 21$, the AGN candidate
surface density is 35 per deg$^2$. For comparison, Stern et al. (2012) finds 61.9$\pm$5.4 per deg$^{-2}$ AGN candidates with 95\%\ reliability for $W2<15.05$, and Assef et al. (in prep) finds 137$\pm$4 per deg$^{-2}$ candidates for $W2<17.11$.

%The $10 \sigma$ $W2$ depth
%of the \wise\, sample corresponds to $W2 = 15.2$.  At this depth,
%the AGN selection criterion, $W1 - W2 > 0.8$, identifies 
%AGN/QSO
%candidates with a surface density of about $56.5\pm0.13$ per deg$^2$.  Considering
%just the optically bright subsample, with $r < 21$, the AGN candidate
%surface density is 35 per deg$^2$.

% FIGURE 8:  W1-W2 vs. W2-W3

% FIGURE 9:  Assef et al. (2010) templates

% FIGURE 10:  galaxy tracks

% FIGURE 11:  SDSS-color-color diagrams

\begin{figure}[!h]
\plotone{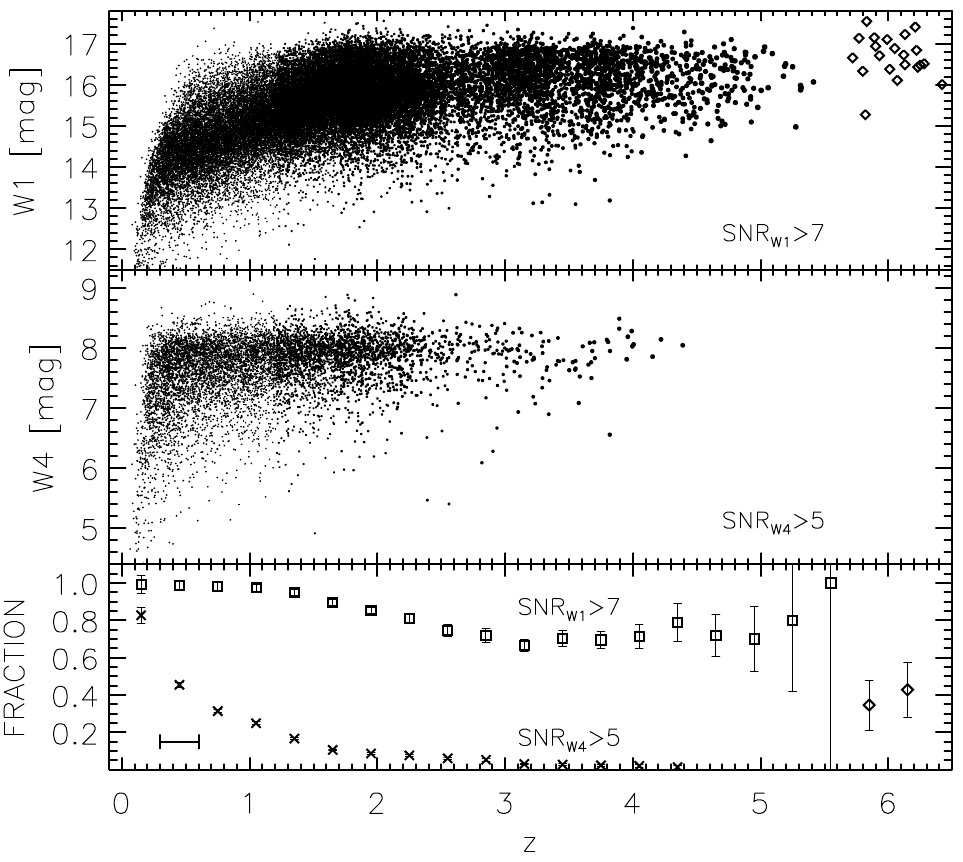}
\caption{The top two panels show the \wise\ magnitudes ($W1$ and $W4$)
 as a function of redshift for the SDSS QSO sample \citep{schneider10}.  As expected, 
more QSOs are detected at 3.4\um\ than at 22\um.  The bottom panel shows the percentages of \wise\ detected SDSS QSOs as a function of $z$.  The $z$\,$>$\,5.7 sources are from \citet{Fan06}, \citet{Jiang09} and CFHT surveys \citep{Willott09, Willott10}. \label{qsoz}}
\end{figure}

SDSS spectroscopic QSOs are optically bright sources. What is the fraction of this population 
detected by \wise\ 3.4 and 22\um? Figure~\ref{qsoz} addresses this question. Overall, \wise\ detects a high fraction of SDSS QSOs from DR7 \citet{schneider10} catalog, and $z>5.7$ QSOs from \citet{Fan06,Jiang09}, \citet{Willott09} and \citet{Willott10}. 
Of the entire QSO sample, only $18.9$\%\ have $W4$ detections at SNR$>5\sigma$, and much higher fraction, 
$89.8$\%\ have $W1$ detections at SNR$>7\sigma$.   We note that \wise\ $W1$ is quite sensitive in detecting 
$z\ge5.8$ QSOs, with a detection rate of $\sim$50\%, as shown in the bottom panel of Figure~\ref{qsoz}. In particular, 
the highest redshift quasar currently known, ULAS~J1120+0461
at $z = 7.085$ \citep{Mortlock11}, is also detected by \wise\ 3.4\um. \citet{Blain12} presents a
detailed discussion of the \wise\, properties  of $z \ge 6$ optically
selected QSOs.

% FIGURE 12 - color-color for qso's

%Figure 13: QSO wise mag vs. z

\subsection{Type-2 AGN candidates \label{sec-type2}}

\begin{figure}[!h]
\plotone{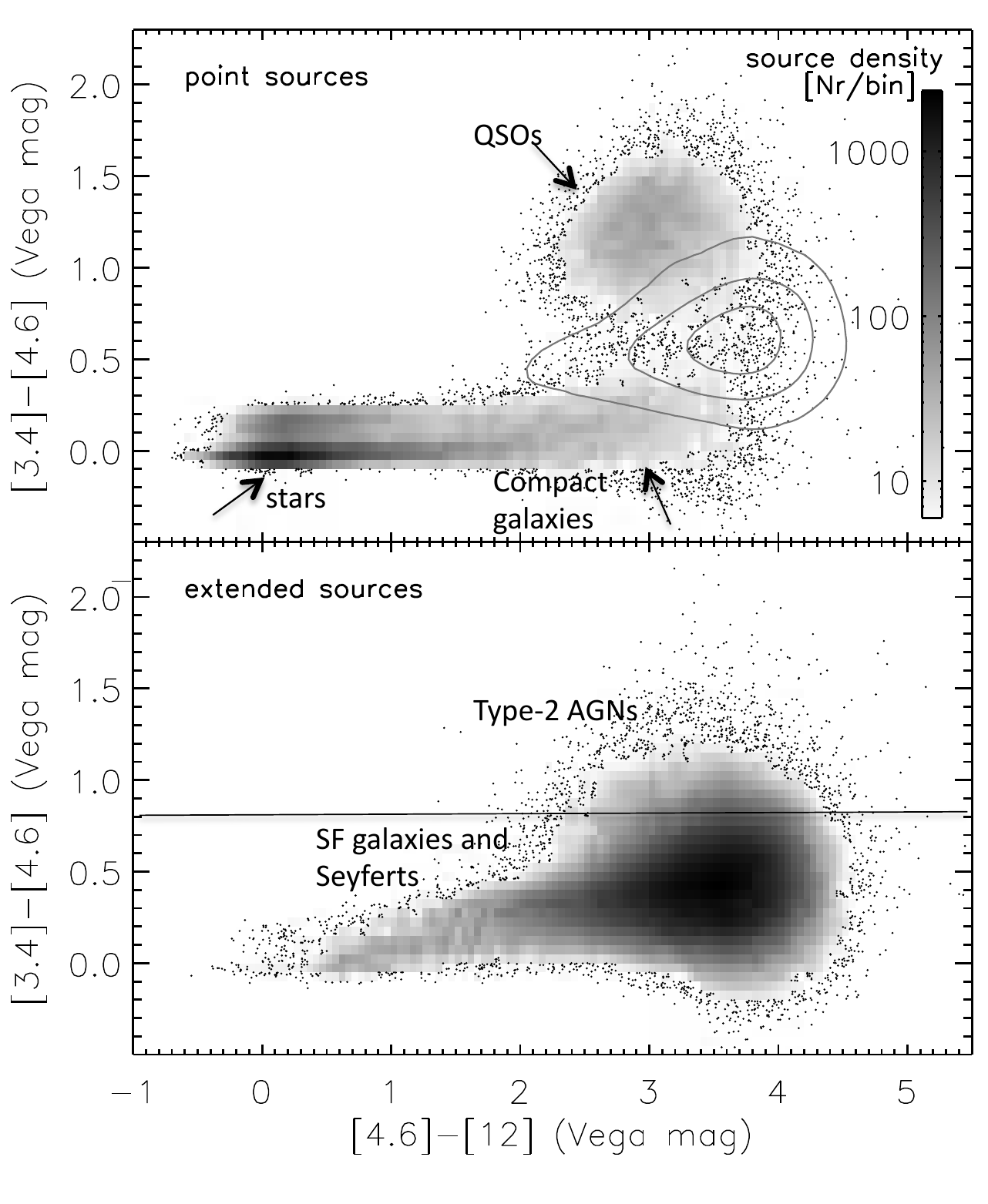}
\caption{\wise\, colors for SDSS unresolved sources (top panel; SDSS
TYPE=6) and extended galaxies (bottom panel; SDSS TYPE=3).  
Comparison of the two panels shows that SDSS point sources have very different $W1-W2$ and $W2-W3$ color distributions from that of extended objects.  In the top panel, 
the solid contours represent the distribution of the extended objects shown below.
Furthermore,
unresolved SDSS sources clearly separate into their two primary
constituencies, stars at the origin and quasars with redder colors.
\label{pointwc}}
\end{figure}

\wise\, is very sensitive to dust-obscured objects. This naturally
leads to two questions: {\it (1)} can \wise\, colors be used to
identify type-2 AGNs; and {\it (2)} what is the relative fraction
of type-1 and type-2 QSO/AGNs selected by \wise?  One simple way to understand
the \wise\ colors of extragalactic sources is to compare mid-IR colors of SDSS unresolved, point
sources (including QSOs, stars and compact galaxies) with that of extended galaxies (including type-2 AGNs and 
normal galaxies).  Figure~\ref{pointwc} illustrates this comparison, showing clear separation between 
strong QSOs and star forming galaxies. Here we take the SDSS photometric catalog, and separate 
sources by their morphological types --- unresolved point sources versus extended galaxies.
This list is then matched with the \wise\ catalog. We do not apply any \wise\ magnitude cut since our purpose is to 
show the difference in mid-IR color distributions of SDSS extended versus pointed sources. 
To ensure reliable star/galaxy classification, we require $r < 21$
\citep[\eg\, 95\%\ reliability quoted in Table 2 in][]{Stoughton02}.   \wise\, colors of
unresolved sources (\eg\, Galactic stars and quasars) are clearly
separated from extended galaxies.  Furthermore, in terms of unresolved
SDSS sources, note that the Galactic stars (at the origin) are
clearly distinct from the unresolved quasars (with redder mid-infrared
colors).

One cautionary note about type-2 AGN is that any discussion and
conclusions will depend how such sources are defined.  In the current
analysis, we define two type-2 AGN samples.  The first sample is the AGN
candidates with $W1 - W2 > 0.8$, $W2 < 15.2$, and with extended SDSS
$r$-band morphologies (SDSS TYPE = 3; $r < 21$).  In general, AGNs have two components, extended 
galaxy and central accreting black holes.
For strong AGNs, the central black hole emission dominates over the extended component in type-1 AGNs, but not in type-2.
Here we adopt the morphology criterion for our selection.
Unfortunately,
this definition only works for optically bright AGN candidates with
spatially resolved host galaxies in the SDSS data. It is also worth noting that 
mid-IR color criteria for AGNs tend to select systems with strong black hole accretions.
For fainter
type-2 AGN candidates, we use an alternative definition: $W1 - W2
> 0.8$, $W2< 15.2$ and very red optical-to-mid-infrared color --- \ie\, optically
faint, \wise\, 4.6\um\ AGN candidates.  As emphasized earlier, when selecting \wise\ AGNs using $W1 - W2$ colors, the $W2<15.2$ criterion is very important
for limiting contamination from high-redshift galaxies to less than 20\% \citep{assef12}. 

\begin{figure}[!htr]
\plotone{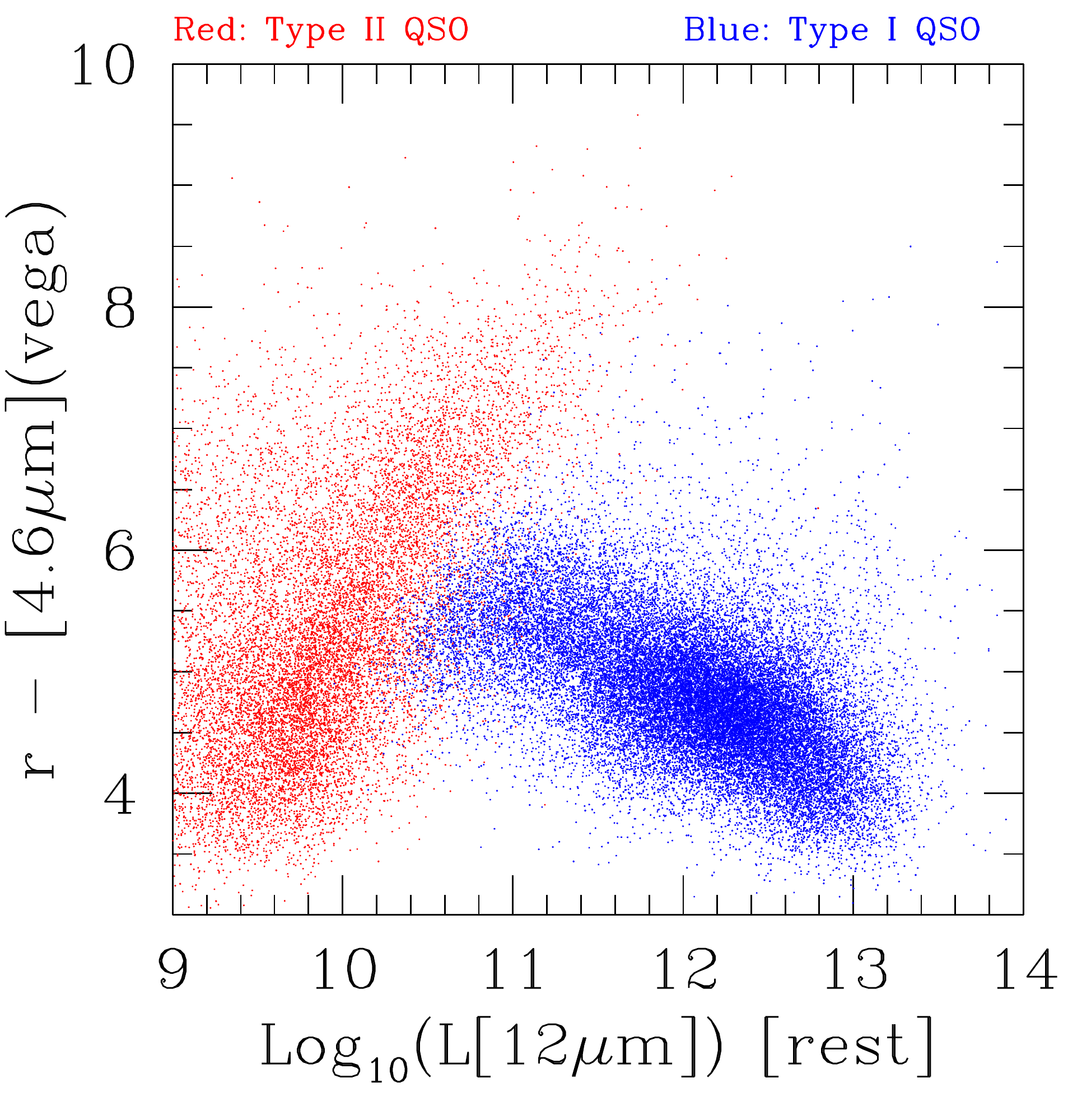}
\caption{
Rest-frame 12\um\ luminosity for the optically bright ($r < 21$)
AGN sample, including small $k$-corrections calculated using
appropriate SEDs for type-1 (blue) and type-2 (red) AGN.  The
differentiation between AGN type is based on optical morphology.
The lack of high-luminosity type-2 AGN is due to the selection effect which
uses SDSS extended morphologies as a criterion. 
More luminous type-2 AGN will be at higher redshifts where the host galaxies brightness falls below the optical magnitude limit imposed
here to ensure reliable morphologies.  Likewise, the host galaxies
of low-luminosity, broad-lined (type-1) AGN are likely detected by
SDSS, and thus such systems are classified as type-2 AGN using our
morphological criterion.
\label{qso12}}
\end{figure}

For the former, optically bright type-2 AGN definition, we estimate $\sim 31\%$ of $r < 21$ \wise-selected AGN/QSOs to be spatially resolved.
This suggests an almost 2:1 type-1 to type-2 AGN ratio.  Since
type-2 AGN here only include bright AGNs with well-resolved host
galaxies in optical images, it is not surprising that the type-2
to type-1 ratio is so low -- \eg\, lower than what is required to
explain the hard X-ray background \citep{gilli07, comastri11}. 
With SDSS + \wise\ data, it is also possible to use $r - W2$ color to select type-2 AGNs.
Figure~\ref{qso12} examines this technique, showing $r - W2$ colors for both type-1 and optically bright type-2 AGNs
with extended SDSS morphologies. The morphological criterion imposes a strong selection function, producing 
an artificial deficit of luminous type-2 AGN in Figure~\ref{qso12}.  Higher luminosity sources will tend to be at higher redshifts and high-redshift obscured AGNs drop below the $r=21$ limit imposed to provide reliable morphologies.
Here $L_{\rm 12\mu m}$ is calculated using spectroscopic or photometric
redshifts, including small $k$-corrections (less than 0.1~magnitude) based on the appropriate
type-1 or type-2 AGN template from \citet{assef10}.  Type-1 and
type-2 AGNs clearly have very different $r - W2$ color distributions, especially at the high luminosity
end.  This is consistent with what has been found by \citet{hickox11}
based on X-ray and \spitzer\, data in the Bo\"otes field.  This result lends support for using $r- W2$ to select type-2 AGNs which do not have reliable optical morphology information.

\begin{figure}[!h]
\plotone{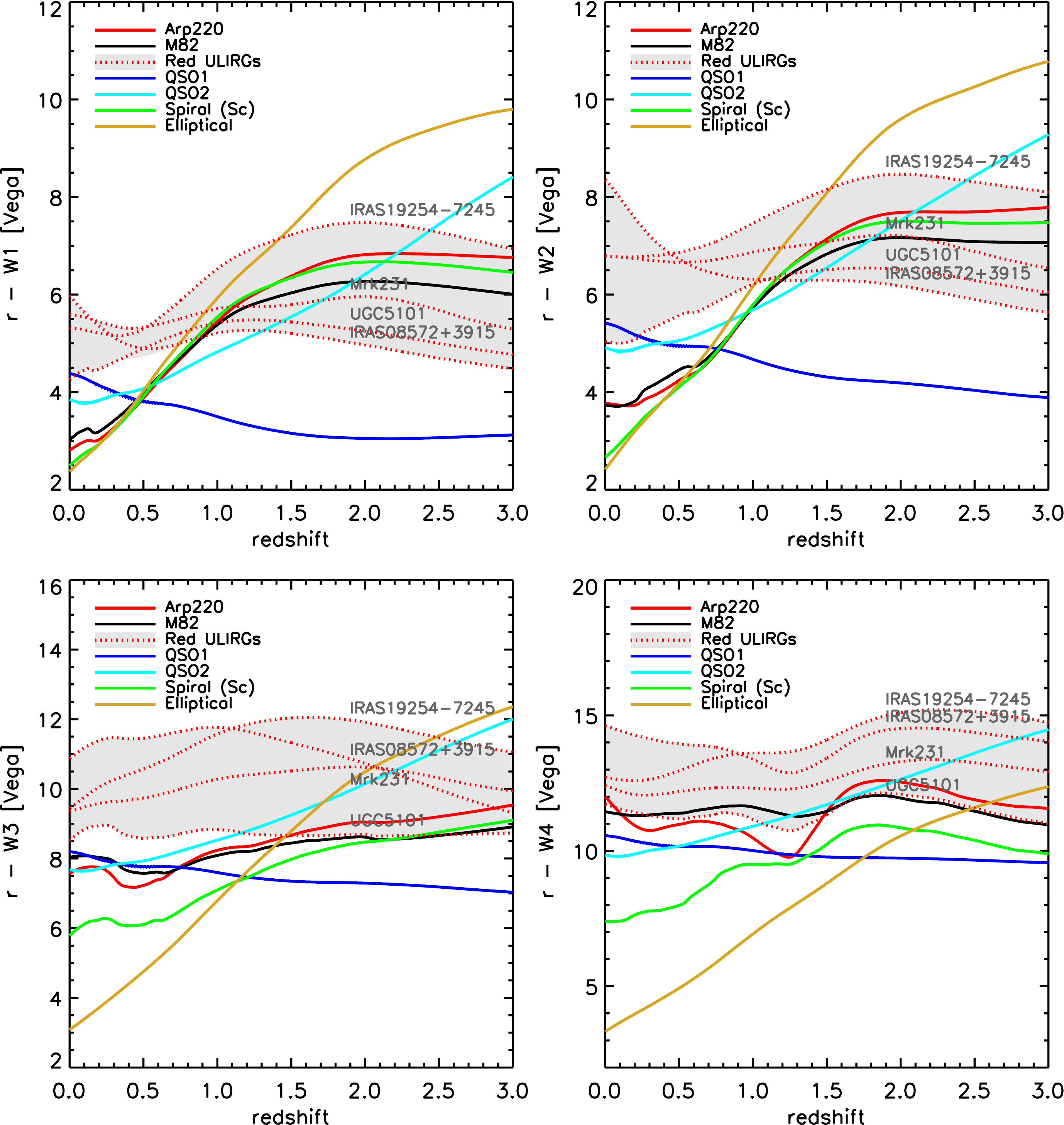}
\caption{Optical-to-\wise\, color as a function of redshift for a
set of galaxy templates.  Red ULIRGs refer to local, highly
dust-obscured ULIRGs with red mid-infrared SEDs, \eg\, Mrk\,231 (a
type-1 AGN), UGC\,5101, IRAS~0872+3915 and IRAS~19254-7245, taken from \citet{polletta07} and \citet{vega08}.
\label{faintr2}}
\end{figure}

\begin{figure}[!htr]
\plotone{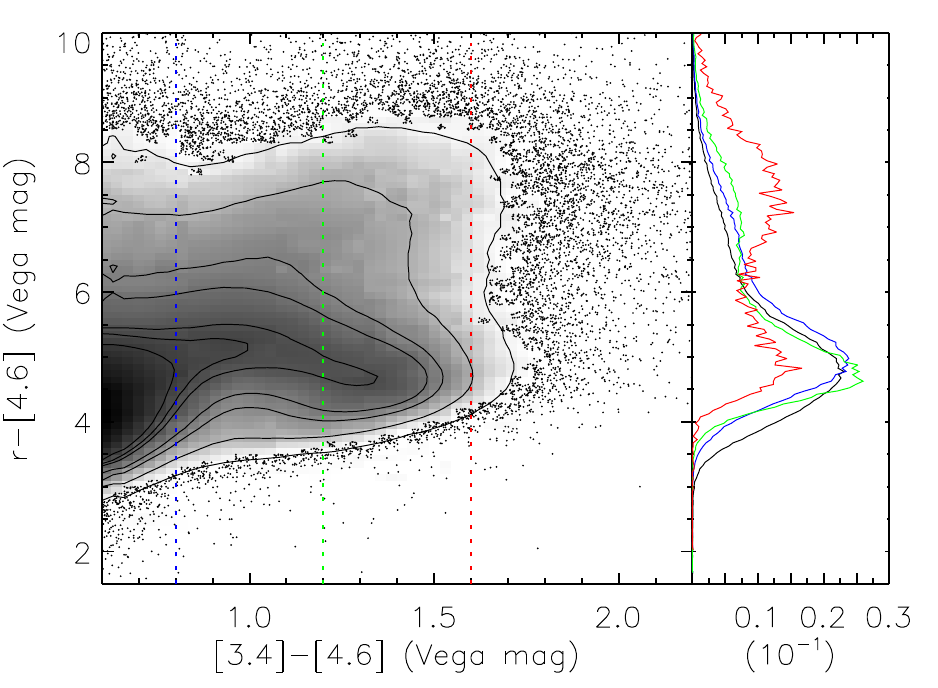}
\caption{Color-color diagram, plotting $r - W2$ vs. $W1 - W2$.
Vertical dashed lines show cuts at various $W1 - W2$ color; the
plot on the right shows the corresponding relative $r - W2$ color
distribution for sources redder than those cuts.  The right-hand side panel is 
the histogram of the  $r-W2$ color distribution; and the x-axis is the number of objects
per $r-W2$ color bin. The black line
shows the full source distribution.   For red \wise\, sources with
$W1 - W2 > 0.8$ (\eg\ AGN candidates), two branches are apparent,
one with blue colors ($r - W2 \sim 4.5$) and a secondary branch
with $r - W2 > 6$.  We propose the latter are type-2
AGN candidates.
\label{faintagn}}
\end{figure}

%For optically faint AGNs lacking reliable optical morphologies, $r
%- W2$ color can be used to identify potential type-2 AGN candidates.
Furthermore, the feasibility of this method is illustrated in 
Figure~\ref{faintr2}, showing the calculated optical/mid-IR colors based on assumed
sets of SED templates as a function of redshift. These galaxy templates are constrained by actual \spitzer\ observations \citep{polletta07,vega08}. We see a strong divergence in optical/mid-IR colors of type-1 and type-2 QSO/AGNs at $z\simgt 0.5$.   Figure~\ref{faintagn}
further examines this technique, showing $r - W2$ vs. $W1 - W2$ for \wise\,
sources with $W2<15.2$, matched with SDSS $r$-band data down to the faintest limits.
The secondary branch in the figure shows that sources
with red $r - W2$ colors are potential type-2 AGNs.  If we use the
criteria of $W1 - W2> 0.8$, $W2 < 15.2$, and $r - W2 > 6$ to define
a sample of red AGN as potential type-2 sources, we find that type-2
candidates account for $\sim 29\%$ of all \wise-selected AGN
candidates to that depth.  The surface density of such a type-2 AGN
candidates is about 16.4 per deg$^2$, whereas $r<21$ bright with resolved morphology, type-2 AGN is about
31\%\ of all AGNs, yielding surface density of 17.3 per deg$^2$.  \citet{stern12}, using the
{\it Hubble Space Telescope} imaging available in the COSMOS field,
find that $\sim 50\%$ of \wise-selected AGN candidates to a depth
of $W2 = 15.2$ are unresolved in $I_{814}$.

Our intention here is to illustrate the methods of using colors to select large samples of type-2 AGN candidates over
wide areas of sky.  Our crude estimate of the type-2 AGN fraction (integrated) is on the order of 1/3rd of all AGNs. However, we caution that to truly understand the implications of these numbers, we will need to have the redshift and luminosity information.  Such studies utilizing \wise\ plus other ancillary spectroscopic data in smaller regions of sky are included in \citet{assef12} and Stern et al. (in prep).

% FIGURE 14 - resolved vs. unresolved in color-color space

% FIGURE 15 - type-1 vs. type-2 AGN  ---   HERE HERE XXXXX

% FIGURE 16 - tracks

% FIGURE 17 - r-W2 vs W1-W2 for AGN

\subsection{High-redshift ULIRG candidates \label{sec-highz}}

One of the primary \wise\, mission goals is to identify extremely
luminous ($L_{IR} \ge 10^{12-13}\, L_\odot$) dusty starbursts and
AGNs (\ie\, ULIRGs and hyper-luminous infrared galaxies, or HyLIRGs)
at high redshift.  We have pursued two basic approaches in finding
high-redshift ULIRGs. One is to use \wise\ data alone, and the
second is to combine \wise\ with optical data.

Using \wise\, colors alone, high-redshift ULIRGs can be identified
by requiring very red, rising mid-infrared SEDs, redder than the
spectral index $\alpha$ $\sim-2.56$ corresponding to the \wise\
sensitivity limits (see \S~\ref{sec-wisedata}). Specifically, this
means significant detections (SNR\,$>$\,3\,-\,5) in $W3$ or $W4$,
but no detections in $W1$ and $W2$, \ie\, so-called $W1W2$-dropouts
\citep{eisenhardt12}.  Follow-up optical spectroscopy  of
more than 100 candidates using the Keck, Gemini, Magellan, and
Palomar telescopes have demonstrated that the majority of these
candidates are indeed at redshifts of 1\,-\,3, implying very high
luminosities given the high fluxes at 12 and 22\um\, \citep{eisenhardt12, jingwen12}.  This type of ULIRG is rare,
with a surface density of just 0.02 sources per deg$^2$.
One interesting feature of these ULIRGs is their extremely steep mid-IR spectral slopes.
Comparing Figure~\ref{modelcc} and \ref{wcc2},
we see that the newly discovered \wise-selected $z$\,$\sim$\,2
ULIRGs have much redder $W2 - W3$ colors than the calculated high-$z$ colors using local ULIRG SED templates, including the reddest one such as IRAS15250+3609. 
%Figure~\ref{wcc2} shows a few examples of these sources at $z > 2$;
%their $W1 - W2$ colors are red, but not well defined, since both
%values are limits.  
\citet{eisenhardt12}, \citet{jingwen12}, \citet{bridge12} and
\citet{yan12} present complete discussions on the $W1W2$-drop
selection, follow-up spectroscopy, as well as far-infrared photometry
from CSO and \herschel\, for some of these extremely luminous
($L_{IR} \sim 10^{13-14}\, L_\odot$) ULIRGs at $z \sim 2$.

\begin{figure}[!htr]
\plotone{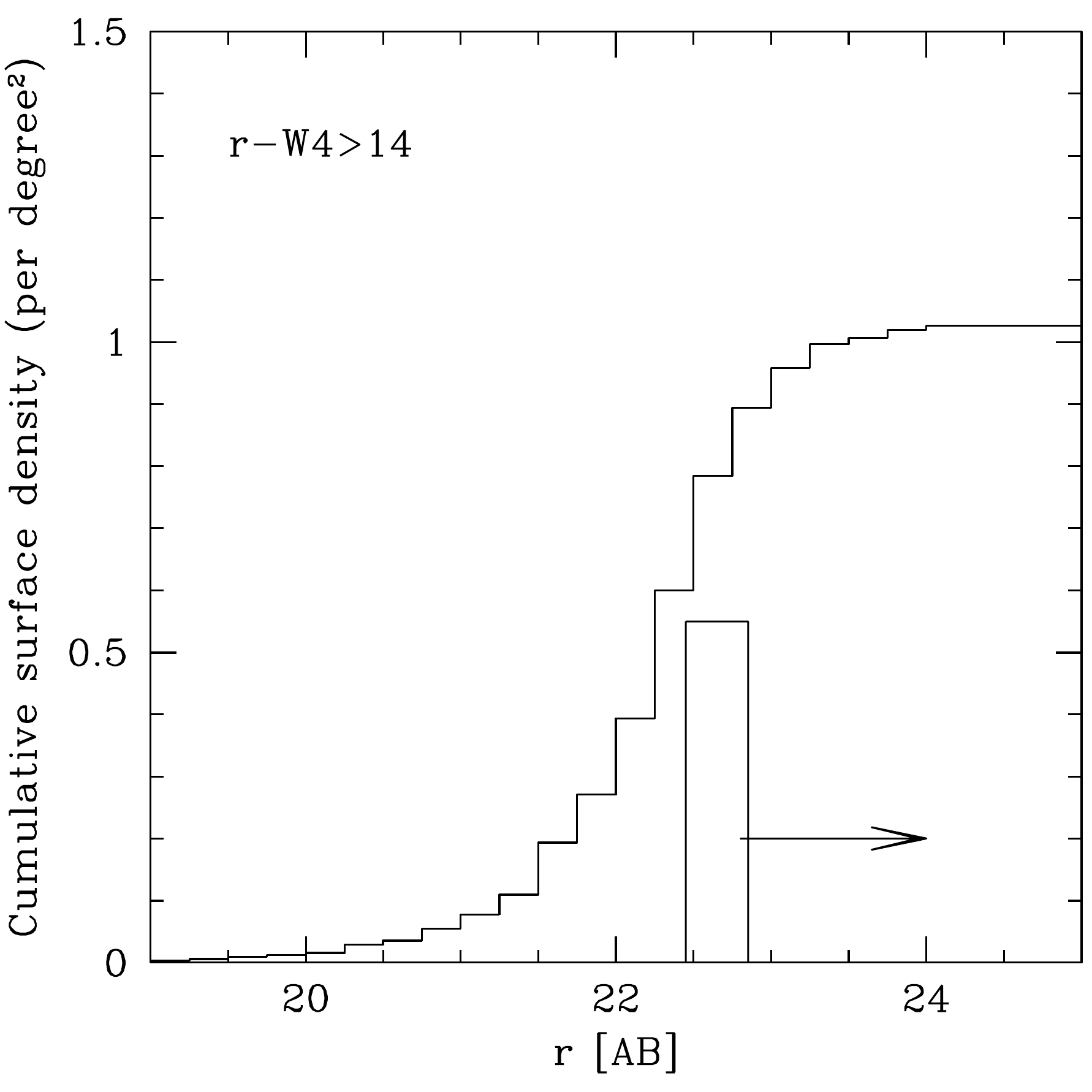}
\caption{Cumulative surface density of high-redshift ULIRG
candidates selected by red optical-to-$W4$ color.  The single bar
with the right side arrow shows sources without SDSS counterparts.
\label{hiz}}
\end{figure}

The second method is to utilize optical/mid-infrared colors. As
shown in Figure~\ref{faintr2}, $r - W1$ color becomes redder at
higher redshift.  This is true as well for the other three \wise\,
bands. Indeed, there is a rich literature using red optical-to-mid-infrared
colors with \spitzer\, to select high-redshift galaxies.  For
example, red 24\um\ to $r$-band colors ($r - [24] > 14$; Vega) have
been used in the Extragalactic First Look Survey (XFLS; \citep{yan04}) and
Bo\"otes field to successfully select many highly obscured galaxies
at $z \sim 2$ \citep{yan05, yan07, dey08}.  As shown in
Figure~\ref{faintr2}, highly dust-reddened local ULIRGs, such as
IRAS~19254-7245, Mrk~231, UGC5101 and IRAS~08572+3915 have very red
colors at any redshifts.  Using $r - W4 > 14$ as the selection
criterion, \wise\, identifies $0.9\pm0.07$ high-redshift
ULIRG candidates per deg$^2$ for SNR$_{W4} > 5$. This signal-to-noise ratio cut 
roughly corresponds to a 22\um\ flux
density $\geq 2.5$~mJy.  This is shown in Figure~\ref{hiz}.  If we instead apply a uniform
flux density cut of $f_\nu(22 \mu m) \ge 5$~mJy, the surface density of ULIRG candidates is $0.41\pm0.05$ per deg$^2$, comparable to that of Dey et al. (2008) using \spitzer\ data over the $\sim$10~deg$^{-2}$ Bo\"otes field with the flux density cut of $f_\nu(24 \mu m)\ge5$~mJy. 
%For comparison with the \spitzer\ study by Dey et al. (2008), we apply a uniform flux cut at $f_\nu (22\mu m)=5$\,mJy, we derived the surface density of $0.41\pm0.05$ per deg$^2$. This is comparable with what is found by Dey et al. (2008) using \spitzer\ data (see their Figure 1; 5 candidates with $f_\nu(24\mu m)>5$\,mJy over 10 sq. degree).  
To verify our analysis, we chose a subset of the SDSS-WISE overlap region covering 180 sq. degrees. All of the red, high-redshift candidates with $r - W4>14$ were visually examined. We found a small percentage of contaminants (8\%).  Our derived surface density values have been corrected for this percentage.
 The \wise\, all-sky data offers an excellent
opportunity to assemble a large sample of 22\um-selected ``dust-obscured
galaxies'', or DOGs, at $z > 1 - 2$.

% FIGURE 18 - r-W4 ULIRGs

% FIGURE 19 -- examples of Keck spectra for red ULIRGs
\begin{figure}[!htr]
\plotone{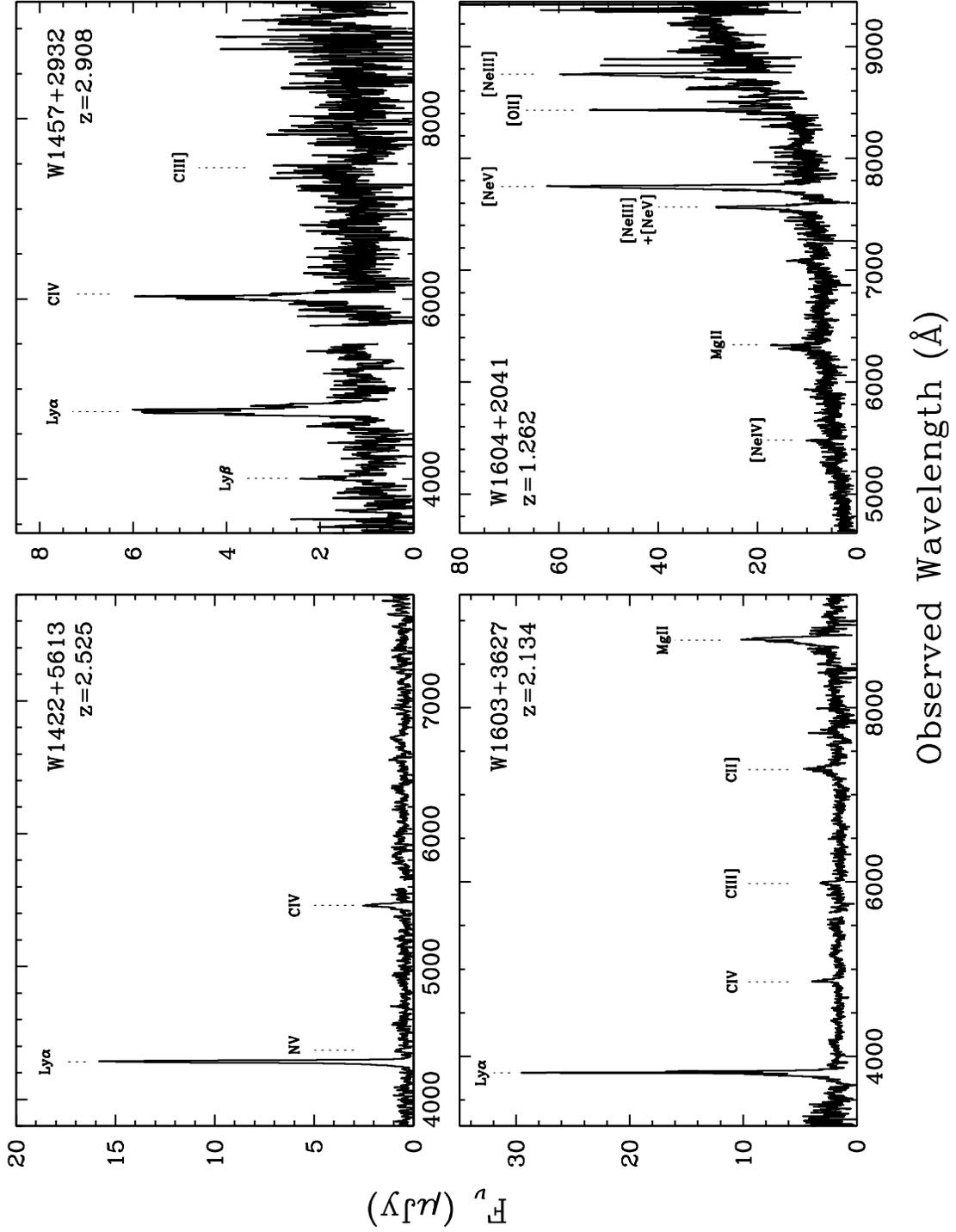}
\caption{Optical spectra of the high-redshift ULIRG candidates selected by $r - W4>14$. \label{exspec}}
\end{figure}

\addtocounter{figure}{-1}
\begin{figure}[!htr]
\plotone{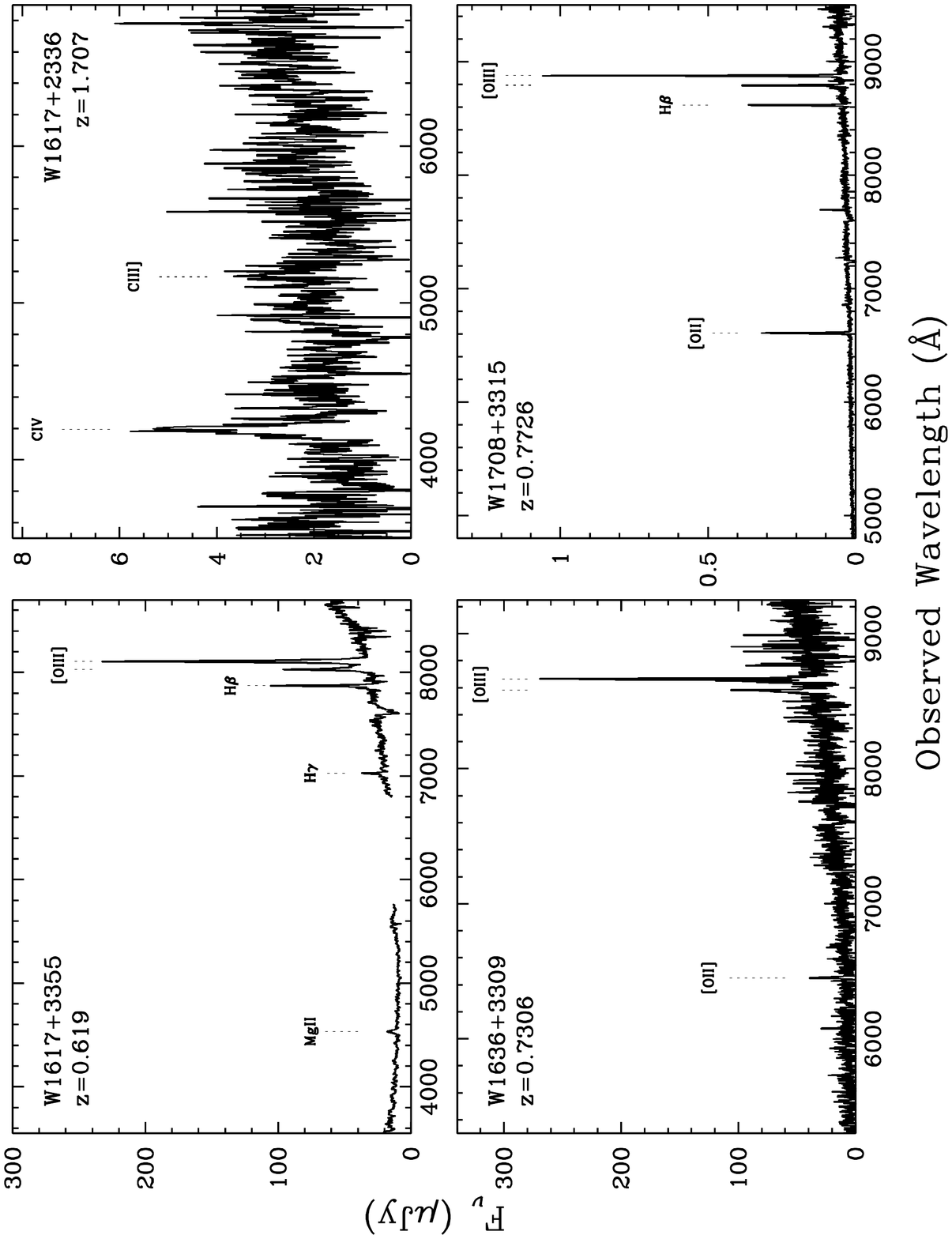}
\caption{Continue.}
\end{figure}

% FIGURE 20 - spec-z colors ULIRGs
\begin{figure}[!htr]
\plotone{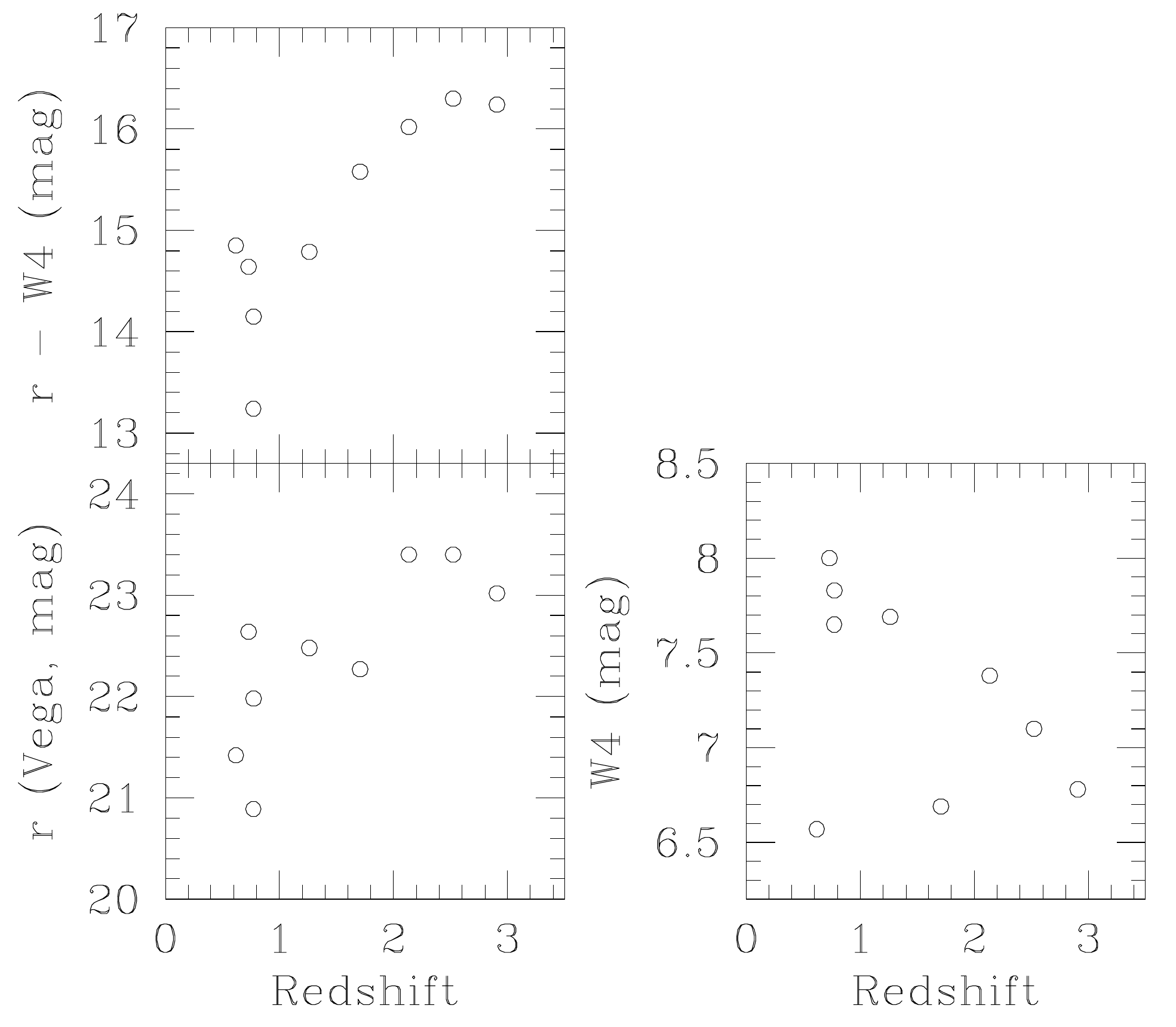}
\caption{The colors and brightness as a function of redshift for the 8 $r-W4>14$ ULIRG candidates with spectroscopic redshifts.  The three panels are intended to investigate trends for selecting high-redshift ULIRG candidates. It is clear that fainter optical magnitudes or redder $r-W4$ colors seem to select ULIRGs at higher redshifts.
\label{speczcolor}}
\end{figure}

\subsection{Keck spectroscopy of a sample of high-redshift ULIRG candidates with very red $r - W4$ color \label{sec-keckspec}}

The \wise\ team has carried out optical spectroscopic follow-up observations of high-redshift ULIRG candidates. These candidates include both $W1W2$-dropouts as well as extremely red optical/mid-IR sources.  \citet{eisenhardt12} focused on one particular $W1W2$-dropout,  J181417.29+341224.9 at $z=2.452$, whose optical spectrum shows a typical star-forming galaxy whereas its IR SED suggests a highly obscured AGN with a bolometric luminosity of $3.7\times10^{13}L_\odot$. \citet{jingwen12} presents optical spectra of a sub-set of $W1W2$-dropouts with millimeter observations, making a comparison between $W1W2$-dropouts and \spitzer-discovered dust obscured galaxies (``DOGs'').  

The follow-up observations included a sample of \wise\ sources with $r - W4 \ge 14$. 
Table~\ref{tab:highzspec} lists the source information  for eight galaxies with good spectra. 
These spectra were taken using Low Resolution Imager and Spectrograph (LRIS,\citep{oke95}) on the Keck telescope
during the nights of March 10, April 10, and May 10, 2011. The total on-target integration times range from 
10 to 15 minutes. Figure~\ref{exspec} shows the wavelength calibrated  spectra in the observed frame for these 8 sources.  Prominent emission lines include \lya, \civ, \mgii, \nev, \oii, and \oiii\  etc, strong spectral features.  W1422+5613 has a \civ\ FWHM of 1480\,km/s, not accounting for the spectral resolution. If we adopt  2000\,km/s for type-1 and type-2 separation based on \civ,  this object is indeed a type-2 AGN.  This cutoff value is very conservative comparing to the mean \civ\ velocity width of 5600\,km/s, estimated the SDSS QSO sample \citep{shen08}. 
Similarly, W1457+2932 is a type-1 AGN with a \civ\ FWHM of 3940\,km/s. W1603+3627 has a \civ\ FWHM of 1790\,km/s, however, its \ciii\ FWHM is broad, 3170\,km/s and the \mgii\ is broad as well. Since \civ\ line is a complex line and could have some absorption. We classify this object as type-1 based on \ciii\ and \mgii\ instead.  W1604+2041 is a type-2 AGN with a \nev\ FWHM of 1350\,km/s. W1617+3355 has a ratio of \oiii\ to H$\beta$ of 3.7, which is in a range that could be either AGN or star-forming based on the 
Baldwin, Phillips \&\ Terlevich (BPT) diagram. Its \oiii\ FWHM is 610\,km/s, making it consistent with either a type-2 AGN or a star-forming galaxies. W1617+2336 is a type-1 AGN with a \civ\ FWHM of 4800\,km/s.  W1636+3309 has a ratio of \oiii\ to H$\beta >$20, which implies an AGN, however, the available spectral line widths for a small number of features are narrow, suggesting it being a type-2.  The classification for this source is not clear. W1708+3315 has [O~III]/H$\beta=3.0$, again making it ambiguous, either a type-2 AGN or a star-forming galaxy.  Table~\ref{tab:highzspec} lists these classifications.
%Based on the velocity width and spectral ratio of these emission lines, we classify these sources in Table~\ref{tab:highzspec}.   The W1603+3627 spectrum detects several high ionization emission lines such as [NeIII], [NeIV] and [NeV], indicating strong nuclear activity.  
Although the number of spectra is small and there are some ambiguous systems,  
the source types among the randomly selected 8 targets are dominated by type-2 AGNs.  This supports the interpretation that  the high mid-IR fluxes at 22\um\  are due to dust emission heated by obscured central AGNs.

We plot the colors and brightness of these 8 sources as a function of their spectroscopic redshifts in Figure~\ref{speczcolor}. It suggests that optically fainter or $r - W4$ redder candidates could be at higher redshifts.  The brightness in the 22\um\ band does not appear to correlate with redshift, with many high-redshift ULIRGs being quite bright in W4.

\section{Discussion and Summary \label{sec-discuss}}

With the public data releases, \wise\, has delivered to the community
the most sensitive all-sky mid-infrared map of our generation.   In this paper,
we present a phenomenological study that characterizes the observational
properties of mid-infrared extragalactic sources and identifies
color selection criteria for isolating large samples of QSOs,
dust-obscured type-2 AGNs and luminous high-redshift ULIRG candidates.

With 5$\sigma$ sensitivities $\le$\,0.05, 0.1, 0.75 and
6\,mJy at 3.4, 4.6, 12 and 22\um, the W1,
W1/2, W1/2/3 and W1/2/3/4 samples have source surface densities
of 8230, 4700, 1235 and 150  per deg$^2$,
respectively.  At the limit of the data, only very red mid-infrared
sources with spectral slopes steeper than $\sim -2.56$ are simultaneously
detected in both $W1$ and $W4$.  The $W1$ source density is one-third
that of the SDSS photometric catalog, suggesting that \wise\, 3.4\um\
does not detect many low-mass, low-luminosity, blue galaxies \citep{emilio12}.
We find that 28\%\ of the $W1$ sample have $r$-band magnitudes
fainter than 22.6, including many sources lacking optical counterparts.
We present observational evidence that suggests that optically faint
3.4\um\ sources are likely early-type galaxies beyond the redshift
limit of the SDSS imaging data.

%  HERE QSO result
%
\wise\ 3.4\um\ data are sensitive to bright quasars. Of the entire SDSS optical 
QSO sample \citep{schneider10}, $89.8$\%\ have $W1$ detections at SNR$>7$.  \wise\ even detects 3.4\um\ emission from the highest, known
redshift QSO, ULAS~J1120+0461
at $z = 7.085$ \citep{Mortlock11}. 
In contrast, only $18.9$\%\ of all SDSS QSO sample have $W4$ detections at SNR$>5$. Some of 
these optically selected bright QSOs detected at 22\um\  are not just bright in all bands, but indeed have strong mid-IR dust emission.

The unique advantage of \wise\, all-sky data for extragalactic
sources is the diagnostic power of its mid-infrared colors.  This
power comes from the  fact that pure stellar
systems, star-forming galaxies and QSO/AGN have distinctly different
near-infrared SEDs (see Figure~\ref{template}),  yielding very different
observed mid-infrared colors for different types of objects. We show that
\wise\, colors alone can separate source populations, including
Galactic stars, star-forming galaxies and QSO/AGN.  We present three
useful applications of \wise\, all-sky data.  {\it (1)} We select
QSO/strong AGN at $z<3$ using $W1 - W2$ color and $W2<15.2$. The magnitude cut
is to limit the contamination from early galaxies at high redshift. This
population of QSOs is interesting to inventory since
a large fraction of QSOs at $z\sim2-3$ are missed by the SDSS optical color
selection.  Beyond $z \sim 3$, the \wise\, QSO color selection
starts to fail as the 3.4\um\ and 4.6\um\ bands sample wavelengths
blue-ward of 1\AA, and the observed \wise\, colors start to have
overlap with low-redshift star-forming galaxies.  {\it (2)} We
demonstrate the possibility for selecting type-2 AGN/QSO candidates
using $W1 - W2 > 0.8$, $W2<15.2$ and $r - W2 > 6$ (or SDSS optically resolved
morphologies for $r < 21$ sources).  The \wise\, data allow a
more complete census of dust-obscured, actively accreting, supermassive
black hole systems than allowed by the current generation of deep
X-ray surveys.  {\it (3)} High-redshift ($z > 2$) dust-obscured,
extremely luminous ULIRGs/HyLIRGs are another population of
extragalactic objects which the \wise\, data can be used to identify.
With follow-up optical spectroscopy, \citet{eisenhardt12}
demonstrates the efficient selection of $z \sim 2-3$ ULIRGs with
extremely red mid-infrared slopes (the $W1W2$-drop-out method).  In
addition, we show that extremely red optical-to-mid-IR colors,
\eg\, using $r - W4>14$, can also be used to select
high-redshift dust-obscured starbursts and AGN. The candidate surface density for this later selection is $0.9\pm0.07$ at
SNR$_{W4}>5$.  This is consistent with the number inferred from the previous studies using \spitzer\ data over much
smaller areas.

Optical spectroscopic follow-up of a small number of $r - W4 >$14
ULIRG candidates confirms that  the color selection indeed works, identifying
IR luminous galaxies over a wide range of redshift, $z\sim 0.7 - 3$.  Despite small 
number statistics, we find indications that optically fainter and $r - W4$ redder sources tend to be
at higher redshifts.  The Keck optical spectra for the 8 sources detect many typical strong emission nebular
lines seen among AGNs, such as \lya, \civ, \nev, \oiii\  lines. 
In addition, of the 8 sources with the optical spectra,  five have  spectral types of 
type-2 AGN, one type-1 AGN with fairly broad \civ\ emission line, and two possible star-forming galaxies.

%The WISE data provides many opportunities for follow-up studies with large number statistical samples. 
%This population of sources are    long sought-after, dust-obscured type-2 AGN candidates using WISE colors, optical morphologies for bright sources and WISE-optical colors.  We demonstrate the possibility for this application using WISE in combination with optical imaging data. The potential type-2 AGN candidates 

%WISE colors alone can provide discriminative power to separate stars (or pure stellar systems), strong AGN/QSOs and star-forming galaxies and weak AGNs.  We show that the color cut [W1\,-\,W2]\,$>$\,0.8 (vega) to select strong AGNs is equivalent to $F_\nu[4.6\mu m]/F_\nu[3.4\mu m]$\,$\ge$\,1.6.  At $z$\,$>$\,3, this color selection does not work, since 4.6, and 3.4\um\ bands move into wavelength shorter than 1\AA, where the QSO SEDs tend to be flatter ({\it i.e.} color bluer). Using the SDSS QSO sample, we illustrate this limitation empirically, and also highlight the unique advantage of WISE colors for selecting QSOs, particularly at $z$\,$\sim$\,2 where SDSS QSOs selection is very incomplete \citep{schneider10}. 

%We illustrate that type I and II QSOs have different [$r$\,-\,W2] color distribution, which could be used as a powerful tool to separate and quantify two populations when combining with photometric redshift estimates. Such measurement would greatly benefit from large number statistics of WISE data.  

\section{Acknowledgement}

%We are grateful to Dominik Riechers for using his Keck time observing object W1604+2041 for 
%us.  
This publication makes use of data products from the {\it Wide-field
Infrared Survey Explorer}, which is a joint project of the University
of California, Los Angeles, and the Jet Propulsion Laboratory/California
Institute of Technology, funded by the National Aeronautics and
Space Administration.  This paper also utilized the publicly available SDSS datasets.
Funding for the SDSS and SDSS-II has been provided by the Alfred P. Sloan Foundation, the Participating Institutions, the National Science Foundation, the U.S. Department of Energy, the National Aeronautics and Space Administration, the Japanese Monbukagakusho, the Max Planck Society, and the Higher Education Funding Council for England. The SDSS Web Site is http://www.sdss.org/.
The SDSS is managed by the Astrophysical Research Consortium for the Participating Institutions. The Participating Institutions are the American Museum of Natural History, Astrophysical Institute Potsdam, University of Basel, University of Cambridge, Case Western Reserve University, University of Chicago, Drexel University, Fermilab, the Institute for Advanced Study, the Japan Participation Group, Johns Hopkins University, the Joint Institute for Nuclear Astrophysics, the Kavli Institute for Particle Astrophysics and Cosmology, the Korean Scientist Group, the Chinese Academy of Sciences (LAMOST), Los Alamos National Laboratory, the Max-Planck-Institute for Astronomy (MPIA), the Max-Planck-Institute for Astrophysics (MPA), New Mexico State University, Ohio State University, University of Pittsburgh, University of Portsmouth, Princeton University, the United States Naval Observatory, and the University of Washington. Some of the data presented herein were obtained at the W.M. Keck Observatory, which is operated as a scientific partnership among the California Institute of Technology, the University of California and the National Aeronautics and Space Administration. The Observatory was made possible by the generous financial support of the W.M. Keck Foundation.  The authors wish to recognize and acknowledge the very significant cultural role and reverence that the summit of Mauna Kea has always had within the indigenous Hawaiian community.  We are most fortunate to have the opportunity to conduct observations from this mountain.

\clearpage

%%%############################################################################
%%%############################################################################
%%% References:

\clearpage

\begin{deluxetable}{cccccccc} 
\rotate
\tabletypesize{\footnotesize}
\setlength{\tabcolsep}{.11cm}
\tablewidth{0pt}
\tablecaption{$r - W4 > 14$, high-$z$ candidates with optical spectra
\label{tab:highzspec}}
\tablehead{ 
\colhead{Source} &
\colhead{$z_{\rm spec}$} &
\colhead{$r$} &
\colhead{$W3$} &
\colhead{$W4$} &
\colhead{$r - W3$} &
\colhead{$r - W4$}  & 
\colhead{type} \\
% units of columns
%\hline \\
 & & mag   &  mag & mag & mag & mag 
}
\startdata
J142228.88+561355.6 &   2.524  &   $>$23.4 &  9.94$\pm0.03$  & 7.10$\pm0.06$ & $>$13.46 & $>$16.30  & type-2 AGN  \\
J145705.91+293231.7 &   2.908  &   23.02   &  9.57$\pm0.03$  & 6.78$\pm0.05$ & 14.45    &  16.24 &  type-1 AGN  \\
J160346.21+362740.2 &   2.134  &   $>$23.4 &  10.86$\pm0.05$ & 7.38$\pm0.07$ & $>$12.54 &$>$16.02  & type-1 AGN  \\
J160456.27+204100.3 &   1.262  &   22.48 &   10.43$\pm0.06$ & 7.69$\pm0.13$  & 12.05    &  14.79  & type-2 AGN \\
J161714.90+335556.4 &   0.619  &   21.42 &   9.52$\ 0.03$  &  6.57$\pm0.05$ & 11.90    &  14.85  & type-2/star forming \\
J161728.91+233646.8 &   1.707  &   22.27 &   9.02$\pm0.03$  & 6.69$\pm0.07$  & 13.25    &  15.58  & type-1 AGN \\
J163619.18+330922.1 &   0.731  &   22.64 &   10.01$\pm0.04$ & 8.00$\pm0.13$  & 12.63    &  14.64  & type-2 AGN \\ 
J170813.71+331538.7 &   0.773  &   21.98 &   9.28$\pm0.03$  & 7.83$\pm0.14$  & 12.70    &  14.15  & type-2/star-forming \\
\enddata
\tablenotetext{1}{All photometry is listed in the Vega system. SDSS $r$ magnitudes have been converted to Vega system by first converting from asinh magnitudes to AB systems, then from AB to Vega. }
\end{deluxetable}

\end{document}